 \documentclass[preprint,review,12pt]{elsarticle}

\usepackage{amssymb}
\usepackage{subcaption}

\usepackage{graphicx} % Allows including images
\usepackage{booktabs} % Allows the use of \toprule, \midrule and \bottomrule in tables
\usepackage{xcolor}
\usepackage{colortbl}

\usepackage{nicefrac}

\usepackage{amsmath}
\usepackage[makeroom]{cancel}
\usepackage[norndcorners,customcolors,nofill]{hf-tikz}
\hfsetbordercolor{red!70}

\usepackage[utf8]{inputenc}
\usepackage[T1]{fontenc}

\usepackage{wasysym}

\usepackage{setspace}

\usepackage{listings}%code snipped

\usepackage{caption}

\usepackage{algorithm}
\usepackage{algpseudocode}

\usepackage{tikz}
\usetikzlibrary{shadows,arrows,shapes.geometric} 
\usetikzlibrary{shapes,fit,patterns,automata}

\usetikzlibrary{quotes,arrows.meta}

\usetikzlibrary{decorations.pathreplacing,calligraphy} %braces

\tikzset{
  annotated cuboid/.pic={
    \tikzset{%
      every edge quotes/.append style={midway, auto},
      /cuboid/.cd,
      #1
    }
    \draw [every edge/.append style={pic actions, densely dashed, opacity=.5}, pic actions]
    (0,0,0) coordinate (o) -- ++(-\cubescale*\cubex,0,0) coordinate (a) -- ++(0,-\cubescale*\cubey,0) coordinate (b) edge coordinate [pos=1] (g) ++(0,0,-\cubescale*\cubez)  -- ++(\cubescale*\cubex,0,0) coordinate (c) -- cycle
    (o) -- ++(0,0,-\cubescale*\cubez) coordinate (d) -- ++(0,-\cubescale*\cubey,0) coordinate (e) edge (g) -- (c) -- cycle
    (o) -- (a) -- ++(0,0,-\cubescale*\cubez) coordinate (f) edge (g) -- (d) -- cycle;
  },
  /cuboid/.search also={/tikz},
  /cuboid/.cd,
  width/.store in=\cubex,
  height/.store in=\cubey,
  depth/.store in=\cubez,
  units/.store in=\cubeunits,
  scale/.store in=\cubescale,
  width=10,
  height=10,
  depth=10,
  units=cm,
  scale=.1,
}

\newif\ifcuboidshade
\newif\ifcuboidemphedge

\tikzset{
  cuboid/.is family,
  cuboid,
  shiftx/.initial=0,
  shifty/.initial=0,
  dimx/.initial=3,
  dimy/.initial=3,
  dimz/.initial=3,
  scale/.initial=1,
  densityx/.initial=1,
  densityy/.initial=1,
  densityz/.initial=1,
  rotation/.initial=0,
  anglex/.initial=0,
  angley/.initial=90,
  anglez/.initial=225,
  scalex/.initial=1,
  scaley/.initial=1,
  scalez/.initial=0.5,
  front/.style={draw=black,fill=white},
  top/.style={draw=black,fill=white},
  right/.style={draw=black,fill=white},
  shade/.is if=cuboidshade,
  shadecolordark/.initial=black,
  shadecolorlight/.initial=white,
  shadeopacity/.initial=0.15,
  shadesamples/.initial=16,
  emphedge/.is if=cuboidemphedge,
  emphstyle/.style={thick},
}

\newcommand{\tikzcuboidkey}[1]{\pgfkeysvalueof{/tikz/cuboid/#1}}

\newcommand{\tikzcuboid}[1]{
    \tikzset{cuboid,#1} % Process Keys passed to command
  \pgfmathsetlengthmacro{\vectorxx}{\tikzcuboidkey{scalex}*cos(\tikzcuboidkey{anglex})*28.452756}
  \pgfmathsetlengthmacro{\vectorxy}{\tikzcuboidkey{scalex}*sin(\tikzcuboidkey{anglex})*28.452756}
  \pgfmathsetlengthmacro{\vectoryx}{\tikzcuboidkey{scaley}*cos(\tikzcuboidkey{angley})*28.452756}
  \pgfmathsetlengthmacro{\vectoryy}{\tikzcuboidkey{scaley}*sin(\tikzcuboidkey{angley})*28.452756}
  \pgfmathsetlengthmacro{\vectorzx}{\tikzcuboidkey{scalez}*cos(\tikzcuboidkey{anglez})*28.452756}
  \pgfmathsetlengthmacro{\vectorzy}{\tikzcuboidkey{scalez}*sin(\tikzcuboidkey{anglez})*28.452756}
  \begin{scope}[xshift=\tikzcuboidkey{shiftx}, yshift=\tikzcuboidkey{shifty}, scale=\tikzcuboidkey{scale}, rotate=\tikzcuboidkey{rotation}, x={(\vectorxx,\vectorxy)}, y={(\vectoryx,\vectoryy)}, z={(\vectorzx,\vectorzy)}]
    \pgfmathsetmacro{\steppingx}{1/\tikzcuboidkey{densityx}}
  \pgfmathsetmacro{\steppingy}{1/\tikzcuboidkey{densityy}}
  \pgfmathsetmacro{\steppingz}{1/\tikzcuboidkey{densityz}}
  \newcommand{\dimx}{\tikzcuboidkey{dimx}}
  \newcommand{\dimy}{\tikzcuboidkey{dimy}}
  \newcommand{\dimz}{\tikzcuboidkey{dimz}}
  \pgfmathsetmacro{\secondx}{2*\steppingx}
  \pgfmathsetmacro{\secondy}{2*\steppingy}
  \pgfmathsetmacro{\secondz}{2*\steppingz}
  \foreach \x in {\steppingx,\secondx,...,\dimx}
  { \foreach \y in {\steppingy,\secondy,...,\dimy}
    {   \pgfmathsetmacro{\lowx}{(\x-\steppingx)}
      \pgfmathsetmacro{\lowy}{(\y-\steppingy)}
      \filldraw[cuboid/front] (\lowx,\lowy,\dimz) -- (\lowx,\y,\dimz) -- (\x,\y,\dimz) -- (\x,\lowy,\dimz) -- cycle;
    }
    }
  \foreach \x in {\steppingx,\secondx,...,\dimx}
  { \foreach \z in {\steppingz,\secondz,...,\dimz}
    {   \pgfmathsetmacro{\lowx}{(\x-\steppingx)}
      \pgfmathsetmacro{\lowz}{(\z-\steppingz)}
      \filldraw[cuboid/top] (\lowx,\dimy,\lowz) -- (\lowx,\dimy,\z) -- (\x,\dimy,\z) -- (\x,\dimy,\lowz) -- cycle;
        }
    }
    \foreach \y in {\steppingy,\secondy,...,\dimy}
  { \foreach \z in {\steppingz,\secondz,...,\dimz}
    {   \pgfmathsetmacro{\lowy}{(\y-\steppingy)}
      \pgfmathsetmacro{\lowz}{(\z-\steppingz)}
      \filldraw[cuboid/right] (\dimx,\lowy,\lowz) -- (\dimx,\lowy,\z) -- (\dimx,\y,\z) -- (\dimx,\y,\lowz) -- cycle;
    }
  }
  \ifcuboidemphedge
    \draw[cuboid/emphstyle] (0,\dimy,0) -- (\dimx,\dimy,0) -- (\dimx,\dimy,\dimz) -- (0,\dimy,\dimz) -- cycle;%
    \draw[cuboid/emphstyle] (0,\dimy,\dimz) -- (0,0,\dimz) -- (\dimx,0,\dimz) -- (\dimx,\dimy,\dimz);%
    \draw[cuboid/emphstyle] (\dimx,\dimy,0) -- (\dimx,0,0) -- (\dimx,0,\dimz);%
    \fi

    \ifcuboidshade
    \pgfmathsetmacro{\cstepx}{\dimx/\tikzcuboidkey{shadesamples}}
    \pgfmathsetmacro{\cstepy}{\dimy/\tikzcuboidkey{shadesamples}}
    \pgfmathsetmacro{\cstepz}{\dimz/\tikzcuboidkey{shadesamples}}
    \foreach \s in {1,...,\tikzcuboidkey{shadesamples}}
    {   \pgfmathsetmacro{\lows}{\s-1}
        \pgfmathsetmacro{\cpercent}{(\lows)/(\tikzcuboidkey{shadesamples}-1)*100}
        \fill[opacity=\tikzcuboidkey{shadeopacity},color=\tikzcuboidkey{shadecolorlight}!\cpercent!\tikzcuboidkey{shadecolordark}] (0,\s*\cstepy,\dimz) -- (\s*\cstepx,\s*\cstepy,\dimz) -- (\s*\cstepx,0,\dimz) -- (\lows*\cstepx,0,\dimz) -- (\lows*\cstepx,\lows*\cstepy,\dimz) -- (0,\lows*\cstepy,\dimz) -- cycle;
        \fill[opacity=\tikzcuboidkey{shadeopacity},color=\tikzcuboidkey{shadecolorlight}!\cpercent!\tikzcuboidkey{shadecolordark}] (0,\dimy,\s*\cstepz) -- (\s*\cstepx,\dimy,\s*\cstepz) -- (\s*\cstepx,\dimy,0) -- (\lows*\cstepx,\dimy,0) -- (\lows*\cstepx,\dimy,\lows*\cstepz) -- (0,\dimy,\lows*\cstepz) -- cycle;
        \fill[opacity=\tikzcuboidkey{shadeopacity},color=\tikzcuboidkey{shadecolorlight}!\cpercent!\tikzcuboidkey{shadecolordark}] (\dimx,0,\s*\cstepz) -- (\dimx,\s*\cstepy,\s*\cstepz) -- (\dimx,\s*\cstepy,0) -- (\dimx,\lows*\cstepy,0) -- (\dimx,\lows*\cstepy,\lows*\cstepz) -- (\dimx,0,\lows*\cstepz) -- cycle;
    }
    \fi 

  \end{scope}
}

\makeatother

\tikzset{cross/.style={cross out, draw=black, minimum size=2*(#1-\pgflinewidth), inner sep=0pt, outer sep=0pt},
cross/.default={1pt}}

\newcounter{bla}

\journal{Computer Physics Communications}

\begin{document}

\begin{frontmatter}



%\title{A \LaTeX{} template for CPC Computer Programs in Physics (CPiP) articles}
\title{A Distributed-memory Tridiagonal Solver \\Based on a Specialised Data Structure \\Optimised for CPU and GPU Architectures}


\author[a]{Semih Akkurt\corref{author}}
\author[b]{Sébastien Lemaire}
\author[b]{Paul Bartholomew}
\author[a]{Sylvain Laizet}
%\author[b]{Third Author}

\cortext[author] {Corresponding author.\\\textit{E-mail address:} s.akkurt18@imperial.ac.uk}
\address[a]{Department of Aeronautics, Imperial College London, SW7 2AZ, United Kingdom}
\address[b]{EPCC, University of Edinburgh, EH8 9BT, United Kingdom}
%\address[b]{Second Address}

\begin{abstract}
Various numerical methods used for solving partial differential equations (PDE) result in tridiagonal systems.
Solving tridiagonal systems on distributed-memory environments is not straightforward, and often requires significant amount of communication.
In this article, we present a novel distributed-memory tridiagonal solver algorithm, DistD2-TDS, based on a specialised data structure.
DistD2-TDS algorithm takes advantage of the diagonal dominance in tridiagonal systems to reduce the communications in distributed-memory environments.
The underlying data structure plays a crucial role for the performance of the algorithm.
First, the data structure improves data localities and makes it possible to minimise data movements via cache blocking and kernel fusion strategies.
Second, data continuity enables a contiguous data access pattern and results in efficient utilisation of the available memory bandwidth. 
Finally, the data layout supports vectorisation on CPUs and thread level parallelisation on GPUs for improved performance.
In order to demonstrate the robustness of the algorithm, we implemented and benchmarked the algorithm on CPUs and GPUs.
We investigated the single rank performance and compared against existing algorithms.
Furthermore, we analysed the strong scaling of the implementation up to 384 NVIDIA H100 GPUs and up to 8192 AMD EPYC 7742 CPUs.
Finally, we demonstrated a practical use case of the algorithm by using compact finite difference schemes to solve a 3D non-linear PDE.
The results demonstrate that DistD2 algorithm can sustain around 66\% of the theoretical peak bandwidth at scale on CPU and GPU based supercomputers.
\end{abstract}

\begin{keyword}
tridiagonal matrix algorithm; distributed tridiagonal matrix solver; compact finite difference schemes
\end{keyword}

\end{frontmatter}

% \linenumbers

\section{Introduction}
\label{sec:intro}
In numerical linear algebra and applied mathematics, a tridiagonal system is a type of matrix that has non-zero entries only on the main diagonal, the diagonal directly above the main, and the diagonal directly below the main. 
Certain numerical discretisations that are used for solving  partial differential equations (PDE) on structured grids in various fields often result in batches of tridiagonal systems that are independent from each other.
For example, discretising the Navier-Stokes equations using high-order compact finite difference schemes \cite{Lele1992} result in batches of tridiagonal systems per each term in the equation.
In addition to compact finite difference schemes, batches of tridiagonal systems also arise when using alternating direction implicit (ADI) solvers \cite{ADI}.

The performance of tridiagonal matrix solvers have been investigated both for single rank and multiple rank strategies including CPU and GPU architectures \cite{Laszlo2016, Balogh2022, Kim2021, SONG2022111443}. 
A typical sub-optimal performance characteristic found in many of these strategies is the reduced performance in solving the $x$-directional tridiagonal systems compared to $y$- and $z$-directional systems.
This is typically due to the Cartesian data structure that is used in these strategies where data points along the $x$-direction are stored next to each other in memory, however there are jumps in memory between data points that are neighbours in the physical space along the $y$-direction or the $z$-direction.
Storing the physically neighbouring entries next to each other in memory eliminates the possibility of certain optimisations due to the sequential operations found in most tridiagonal matrix algorithms. 
In particular, on CPU architectures vectorisation is harder to achieve due to the dependencies between neighbouring elements, and on GPU architectures it becomes harder to utilise all the threads in a thread block efficiently. 
Laszlo et al. \cite{Laszlo2016} proposed a methodology to solve this problem based on carrying out local transposes in CPU cache or GPU shared memory. Although this strategy performs significantly better than a naive approach where the discrepancy in performance between $x$- and $y$-directional operations can be up to 8x, there is still a meaningful discrepancy of up to 2x especially when the domain size is 512 and above on GPUs.

In this paper, we propose a specialised data structure and develop a novel algorithm for improving performance of tridiagonal matrix solvers both on CPU and GPU architectures.
Most importantly, the proposed data structure enables a linear data access pattern, makes it easy to vectorise the operations, and improves data locality which in turn results in a better cache utilisation regardless of the spatial direction of the tridiagonal systems. 
Moreover, we also demonstrate the suitability of this new data structure for kernel fusion strategies, where the mathematical operations are combined at source code level for reducing data movement requirements and improving performance.
This strategy is essential for performance for bandwidth bound algorithms especially on GPU architectures due to their relatively smaller cache memory compared to CPUs.

Furthermore, as the domain where these batches of tridiagonal systems are solved are often quite large, ranging from a billion degree-of-freedom (DOF) to up to a few trillion DOF, employing efficient parallel strategies that can leverage large scale supercomputers is very important.
Therefore, we extend our novel strategy to include a distributed-memory tridiagonal matrix algorithm named DistD2-TDS. %as well.
For simplicity, we focus on diagonally dominant tridiagonal systems for the distributed-memory strategy, taking advantage of the simplified communication pattern of such systems, as discussed in \cite{Sun1995}.
In order to provide real world use cases of diagonally dominant tridiagonal systems, we use higher-order compact finite difference schemes, also known as spatially implicit schemes, to discretise PDEs of interest and apply the proposed algorithm.
It should be noted that the proposed strategy can be extended for solving generic tridiagonal systems as well. 
Further details regarding the limitations of our strategy and potential extensions are discussed in detail in Section 4.

The remainder of the paper is structured as follows. In section \ref{sec:TDMA} we discuss tridiagonal matrix algorithms including serial and distributed-memory strategies.
Section \ref{sec:perf} discusses performance considerations focusing on operations typically found in tridiagonal matrix algorithms. %particularly focusing on the forward and backwards sweeps found in many tridiagonal matrix algorithms. 
Then we propose our novel strategy in Section \ref{sec:method}, describing the DistD2-TDS algorithm in detail. 
Section \ref{sec:results} presents performance results and also provides a practical application of the proposed strategy on a 3D non-linear PDE. Finally, conclusions %and future work 
are discussed in Section \ref{sec:conclusion}.

\section{Tridiagonal Matrix Algorithms}\label{sec:TDMA}
Tridiagonal systems can be described in matrix form $\mathbf{A}\mathbf{u} = \mathbf{d}$ and can be expanded as
%Tridiagonal systems can be shown in matrix form $A\cdot \mathbf{u} = \mathbf{d}$ as
\begin{equation}\label{eq:TDS}
 \begin{bmatrix}
   b_1 & c_1 &  & \\ 
   a_2 & b_2 & c_2 & \\ 
       & a_3 & b_3 & c_3\\
       &     & \ddots & \ddots & \ddots \\ 
&       &     &    a_{n-1}    &  b_{n-1}  & c_{n-1} \\
&       &     &     & a_n & b_n 
 \end{bmatrix}
 \begin{bmatrix}
 u_1 \\ u_2 \\ u_3 \\ \vdots \\ u_{n-1}\\ u_{n}
 \end{bmatrix}
=
 \begin{bmatrix}
 d_1 \\ d_2 \\ d_3 \\ \vdots \\ d_{n-1} \\ d_{n}
 \end{bmatrix}.
\end{equation}
The coefficient matrix $\mathbf{A}$ consist of 3 bands where we refer $a_i$ as the lower diagonal, $b_i$ as the diagonal, and $c_i$ as the upper diagonal coefficients, $u_i$ is the solution vector, $d_i$ is the right hand side (RHS) vector, and $n$ is the size of the tridiagonal system.
The banded structure in tridiagonal matrices makes it possible to use simplified algorithms to solve these systems.
In the remainder of this section, we investigate existing algorithms for solving tridiagonal systems.
We particularly focus on the data movement requirements for these algorithms when analysing their performance limitations, as it is the main bottleneck due to the low arithmetic intensity of these algorithms. 

\subsection{Serial Algorithm - Thomas Algorithm}
The Thomas algorithm \cite{THOMAS} is effectively a simplified Gauss elimination, consisting of forward and backward passes to eliminate the lower and upper diagonals in the coefficient matrix $\mathbf{A}$ as described in Algorithm \ref{alg:thom}.

\begin{singlespacing}
\begin{algorithm}
\caption{Thomas algorithm.}\label{alg:thom}
\begin{algorithmic}[1]
\State $u_1 \gets d_1/b_1$
\State $c_1 \gets c_1/b_1$
\For{$i = 2:n$} \Comment{Forward pass}
  \State $w \gets 1/(b_i - a_i c_{i-1})$
  \State $u_i \gets w(d_i - a_i u_{i-1})$  %\Comment{This is a comment}
  \State $c_i \gets wc_i$
\EndFor
\For{$i = n-1:1$} \Comment{Backward pass}
  \State $u_i \gets u_i - c_iu_{i+1}$
\EndFor
\end{algorithmic}
\end{algorithm}
\end{singlespacing}

\begin{figure}[h!]
  \centering
\begin{subfigure}{0.31\textwidth}
\begin{tikzpicture}[scale=0.28]
%\draw (0,0) grid (15,15);
\draw[step=1.0, lightgray] (0,0) grid (15, 15);
\draw[step=15.0, thick] (0,0) grid (15, 15);
%\draw[very thick, step=5.0] (0.1,0.1) grid (14.9, 14.9);
%\draw[xshift=10] (15,0) grid [xstep=1,ystep=5] (16, 15);
%\draw[xshift=10] (17,0) grid [xstep=1,ystep=5] (18, 15);
%\draw[black, very thick] (-0.4, 5) -- (15.4, 5);
%\draw[thick] (-0.4, 10) -- (15.4, 10);
\foreach \i in {1, 2, ..., 15}{
\filldraw[darkgray] (\i-0.5,15.5-\i) circle (8pt);
}
\foreach \i in {1, 2, ..., 14}{
\filldraw[darkgray] (\i-0.5,14.5-\i) circle (8pt);
\filldraw[darkgray] (\i+0.5,15.5-\i) circle (8pt);
}
\end{tikzpicture}
\caption{Tridiagonal system} \label{fig:thom_a}
\end{subfigure}
\begin{subfigure}{0.31\textwidth}
\begin{tikzpicture}[scale=0.28]
%\draw (0,0) grid (15,15);
\draw[step=1.0, lightgray] (0,0) grid (15, 15);
\draw[step=15.0, thick] (0,0) grid (15, 15);
%\draw[very thick, step=5.0] (0.1,0.1) grid (14.9, 14.9);
%\draw[xshift=10] (15,0) grid [xstep=1,ystep=5] (16, 15);
%\draw[xshift=10] (17,0) grid [xstep=1,ystep=5] (18, 15);
%\draw[black, very thick] (-0.4, 5) -- (15.4, 5);%(18.8, 5);
%\draw[thick] (-0.4, 10) -- (15.4, 10);
\foreach \i in {1, 2, ..., 15}{
\filldraw[darkgray] (\i-0.5,15.5-\i) circle (8pt);
}
\foreach \i in {1, 2, ..., 14}{
%\filldraw[darkgray] (\i-0.5,14.5-\i) circle (8pt);
\filldraw[darkgray] (\i+0.5,15.5-\i) circle (8pt);
}
\end{tikzpicture}
\caption{After forward pass} \label{fig:thom_b}
\end{subfigure}
\begin{subfigure}{0.31\textwidth}
\begin{tikzpicture}[scale=0.28]
%\draw (0,0) grid (15,15);
\draw[step=1.0, lightgray] (0,0) grid (15, 15);
\draw[step=15.0, thick] (0,0) grid (15, 15);
%\draw[very thick, step=5.0] (0.1,0.1) grid (14.9, 14.9);
%\draw[xshift=10] (15,0) grid [xstep=1,ystep=5] (16, 15);
%\draw[xshift=10] (17,0) grid [xstep=1,ystep=5] (18, 15);
%\draw[black, very thick] (-0.4, 5) -- (15.4, 5);%(18.8, 5);
%\draw[thick] (-0.4, 10) -- (15.4, 10);
\foreach \i in {1, 2, ..., 15}{
\filldraw[darkgray] (\i-0.5,15.5-\i) circle (8pt);
}
\end{tikzpicture}
\caption{After backward pass} \label{fig:thom_c}
\end{subfigure}
  \caption{Thomas algorithm for solving tridiagonal matrices. Non-zero entries in the banded matrix are represented by gray circles. Forward pass eliminates the lower diagonal, and backward pass eliminates the upper diagonal. RHS and the unknown vectors are omitted for simplicity.}
  \label{fig:thom}
\end{figure}
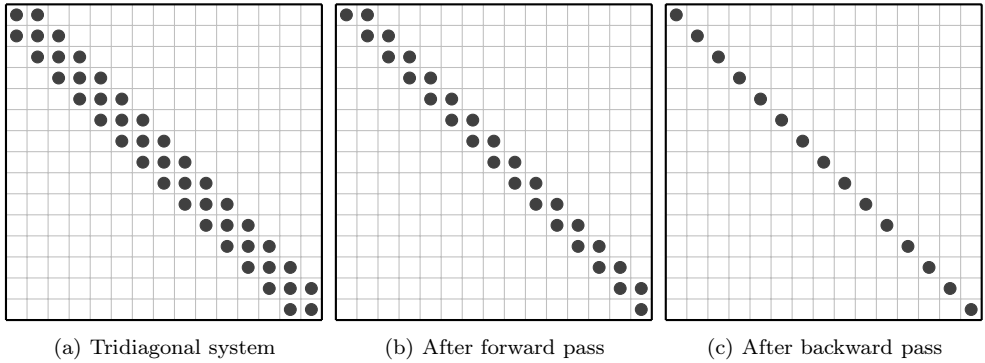

\begin{figure}[h!]
  \centering
\begin{subfigure}{0.31\textwidth}
\scalebox{0.5}{
\begin{tikzpicture}
    \tikzcuboid{%
    shiftx=0cm,%
    shifty=0cm,%
    scale=1,%
    rotation=0,%
    densityx=2,%
    densityy=2,%
    densityz=2,%
    dimx=5,%
    dimy=1,%
    dimz=5,%
    scalex=0.95,%
    scaley=1,%
    scalez=1,%
    anglex=0,%
    angley=90,%
    anglez=225,%
    front/.style={draw=red!50!black,fill=red!50!white},%
    top/.style={draw=red!50!black,fill=red!50!white},%
    right/.style={draw=red!50!black,fill=red!50!white},%
    emphedge=false,%
    }
\begin{scope}[shift={(0,1.5,0)}]
    \tikzcuboid{%
    front/.style={draw=green!50!black,fill=green!50!white},%
    top/.style={draw=green!50!black,fill=green!50!white},%
    right/.style={draw=green!50!black,fill=green!50!white},%
    };   
\end{scope}
\begin{scope}[shift={(0,3,0)}]
    \tikzcuboid{%
    front/.style={draw=blue!50!black,fill=blue!50!white},%
    top/.style={draw=blue!50!black,fill=blue!50!white},%
    right/.style={draw=blue!50!black,fill=blue!50!white},%
    }   
\end{scope}
\begin{scope}[shift={(0,4.5,0)}]
    \tikzcuboid{%
    front/.style={draw=yellow!50!black,fill=yellow!50!white},%
    top/.style={draw=yellow!50!black,fill=yellow!50!white},%
    right/.style={draw=yellow!50!black,fill=yellow!50!white},%
    }   
\end{scope}
    \draw[thick,->] (-2.5,4.5,0) -- (-1.5,4.5,0) node[anchor=north east, font=\Large]{$x$};
    \draw[thick,->] (-2.5,4.5,0) -- (-2.5,5.5,0) node[anchor=north west, font=\Large]{$y$};
    \draw[thick,->] (-2.5,4.5,0) -- (-2.5,4.5,1) node[anchor=east, font=\Large]{$z$};
\end{tikzpicture}
}
\caption{1D Decomposition} \label{fig:m_a}
\end{subfigure}
\begin{subfigure}{0.31\textwidth}
\begin{tikzpicture}
    \tikzcuboid{%
    shiftx=0cm,%
    shifty=0cm,%
    scale=0.5,%
    rotation=0,%
    densityx=2,%
    densityy=2,%
    densityz=2,%
    dimx=5,%
    dimy=5.5,%
    dimz=1,%
    scalex=1,%
    scaley=1,%
    scalez=1,%
    anglex=0,%
    angley=90,%
    anglez=225,%
    front/.style={draw=red!50!black,fill=red!50!white},%
    top/.style={draw=red!50!black,fill=red!50!white},%
    right/.style={draw=red!50!black,fill=red!50!white},%
    emphedge=false,%
    }
\begin{scope}[shift={(0,0,1.2)}]
    \tikzcuboid{%
    front/.style={draw=green!50!black,fill=green!50!white},%
    top/.style={draw=green!50!black,fill=green!50!white},%
    right/.style={draw=green!50!black,fill=green!50!white},%
    };   
\end{scope}
\begin{scope}[shift={(0,0,2.4)}]
    \tikzcuboid{%
    front/.style={draw=blue!50!black,fill=blue!50!white},%
    top/.style={draw=blue!50!black,fill=blue!50!white},%
    right/.style={draw=blue!50!black,fill=blue!50!white},%
    }   
\end{scope}
\begin{scope}[shift={(0,0,3.6)}]
    \tikzcuboid{%
    front/.style={draw=yellow!50!black,fill=yellow!50!white},%
    top/.style={draw=yellow!50!black,fill=yellow!50!white},%
    right/.style={draw=yellow!50!black,fill=yellow!50!white},%
    }   
\end{scope}
\end{tikzpicture}
\caption{1D Decomposition} \label{fig:m_b}
\end{subfigure}
\begin{subfigure}{0.31\textwidth}
\scalebox{0.5}{
\begin{tikzpicture}
    \tikzcuboid{%
    shiftx=0cm,%
    shifty=0cm,%
    scale=1,%
    rotation=0,%
    densityx=2,%
    densityy=2,%
    densityz=2,%
    dimx=3,%
    dimy=3,%
    dimz=1,%
    scalex=1,%
    scaley=1,%
    scalez=1,%
    anglex=0,%
    angley=90,%
    anglez=225,%
    front/.style={draw=gray!50!black,fill=gray!50!white},%
    top/.style={draw=gray!50!black,fill=gray!50!white},%
    right/.style={draw=gray!50!black,fill=gray!50!white},%
    emphedge=false,%
    }
\begin{scope}[shift={(0,4,0)}]
    \tikzcuboid{%
    front/.style={draw=green!50!black,fill=green!50!white},%
    top/.style={draw=green!50!black,fill=green!50!white},%
    right/.style={draw=green!50!black,fill=green!50!white},%
    };   
\end{scope}
\begin{scope}[shift={(4,0,0)}]
    \tikzcuboid{%
    front/.style={draw=red!50!black,fill=red!50!white},%
    top/.style={draw=red!50!black,fill=red!50!white},%
    right/.style={draw=red!50!black,fill=red!50!white},%
    }   
\end{scope}
\begin{scope}[shift={(4,4,0)}]
    \tikzcuboid{%
    front/.style={draw=yellow!50!black,fill=yellow!50!white},%
    top/.style={draw=yellow!50!black,fill=yellow!50!white},%
    right/.style={draw=yellow!50!black,fill=yellow!50!white},%
    }   
\end{scope}

\begin{scope}[shift={(0,0,3)}]
    \tikzcuboid{%
    front/.style={draw=cyan!50!black,fill=cyan!50!white},%
    top/.style={draw=cyan!50!black,fill=cyan!50!white},%
    right/.style={draw=cyan!50!black,fill=cyan!50!white},%
    };   
\end{scope}
\begin{scope}[shift={(0,4,3)}]
    \tikzcuboid{%
    front/.style={draw=orange!50!black,fill=orange!50!white},%
    top/.style={draw=orange!50!black,fill=orange!50!white},%
    right/.style={draw=orange!50!black,fill=orange!50!white},%
    };   
\end{scope}
\begin{scope}[shift={(4,0,3)}]
    \tikzcuboid{%
    front/.style={draw=lime!50!black,fill=lime!50!white},%
    top/.style={draw=lime!50!black,fill=lime!50!white},%
    right/.style={draw=lime!50!black,fill=lime!50!white},%
    }   
\end{scope}
\begin{scope}[shift={(4,4,3)}]
    \tikzcuboid{%
    front/.style={draw=blue!50!black,fill=blue!50!white},%
    top/.style={draw=blue!50!black,fill=blue!50!white},%
    right/.style={draw=blue!50!black,fill=blue!50!white},%
    }   
\end{scope}
\end{tikzpicture}
}
\caption{3D Decomposition} \label{fig:m_c}
\end{subfigure}

\begin{subfigure}{0.31\textwidth}
\scalebox{0.5}{
\begin{tikzpicture}
    \tikzcuboid{%
    shiftx=0cm,%
    shifty=0cm,%
    scale=1,%
    rotation=0,%
    densityx=2,%
    densityy=2,%
    densityz=2,%
    dimx=1,%
    dimy=1,%
    dimz=5,%
    scalex=1,%
    scaley=1,%
    scalez=0.8,%
    anglex=0,%
    angley=90,%
    anglez=225,%
    front/.style={draw=red!50!black,fill=red!50!white},%
    top/.style={draw=red!50!black,fill=red!50!white},%
    right/.style={draw=red!50!black,fill=red!50!white},%
    emphedge=false,%
    }
\begin{scope}[shift={(0,1.5,0)}]
    \tikzcuboid{%
    front/.style={draw=green!50!black,fill=green!50!white},%
    top/.style={draw=green!50!black,fill=green!50!white},%
    right/.style={draw=green!50!black,fill=green!50!white},%
    };   
\end{scope}
\begin{scope}[shift={(0,3,0)}]
    \tikzcuboid{%
    front/.style={draw=blue!50!black,fill=blue!50!white},%
    top/.style={draw=blue!50!black,fill=blue!50!white},%
    right/.style={draw=blue!50!black,fill=blue!50!white},%
    }   
\end{scope}
\begin{scope}[shift={(0,4.5,0)}]
    \tikzcuboid{%
    front/.style={draw=yellow!50!black,fill=yellow!50!white},%
    top/.style={draw=yellow!50!black,fill=yellow!50!white},%
    right/.style={draw=yellow!50!black,fill=yellow!50!white},%
    }   
\end{scope}

\begin{scope}[shift={(1.5,0,0)}]
    \tikzcuboid{%
    front/.style={draw=lime!50!black,fill=lime!50!white},%
    top/.style={draw=lime!50!black,fill=lime!50!white},%
    right/.style={draw=lime!50!black,fill=lime!50!white},%
    };   
\end{scope}
\begin{scope}[shift={(1.5,1.5,0)}]
    \tikzcuboid{%
    front/.style={draw=magenta!50!black,fill=magenta!50!white},%
    top/.style={draw=magenta!50!black,fill=magenta!50!white},%
    right/.style={draw=magenta!50!black,fill=magenta!50!white},%
    };   
\end{scope}
\begin{scope}[shift={(1.5,3,0)}]
    \tikzcuboid{%
    front/.style={draw=olive!50!black,fill=olive!50!white},%
    top/.style={draw=olive!50!black,fill=olive!50!white},%
    right/.style={draw=olive!50!black,fill=olive!50!white},%
    }   
\end{scope}
\begin{scope}[shift={(1.5,4.5,0)}]
    \tikzcuboid{%
    front/.style={draw=pink!50!black,fill=pink!50!white},%
    top/.style={draw=pink!50!black,fill=pink!50!white},%
    right/.style={draw=pink!50!black,fill=pink!50!white},%
    }  % #brown%teal%violet%lime%pink%purple
\end{scope}

\begin{scope}[shift={(3,0,0)}]
    \tikzcuboid{%
    front/.style={draw=white!50!black,fill=white!50!white},%
    top/.style={draw=white!50!black,fill=white!50!white},%
    right/.style={draw=white!50!black,fill=white!50!white},%
    };   
\end{scope}
\begin{scope}[shift={(3,1.5,0)}]
    \tikzcuboid{%
    front/.style={draw=orange!50!black,fill=orange!50!white},%
    top/.style={draw=orange!50!black,fill=orange!50!white},%
    right/.style={draw=orange!50!black,fill=orange!50!white},%
    };    %purple
\end{scope}
\begin{scope}[shift={(3,3,0)}]
    \tikzcuboid{%
    front/.style={draw=gray!50!black,fill=gray!50!white},%
    top/.style={draw=gray!50!black,fill=gray!50!white},%
    right/.style={draw=gray!50!black,fill=gray!50!white},%
    }   
\end{scope}
\begin{scope}[shift={(3,4.5,0)}]
    \tikzcuboid{%
    front/.style={draw=cyan!50!black,fill=cyan!50!white},%
    top/.style={draw=cyan!50!black,fill=cyan!50!white},%
    right/.style={draw=cyan!50!black,fill=cyan!50!white},%
    }   
\end{scope}
\end{tikzpicture}
}
\caption{2D Decomposition} \label{fig:m_d}
\end{subfigure}
\begin{subfigure}{0.31\textwidth}
\scalebox{0.5}{
\begin{tikzpicture}
    \tikzcuboid{%
    shiftx=0cm,%
    shifty=0cm,%
    scale=1,%
    rotation=0,%
    densityx=2,%
    densityy=2,%
    densityz=2,%
    dimx=1,%
    dimy=5.5,%
    dimz=1,%
    scalex=1,%
    scaley=1,%
    scalez=0.75,%
    anglex=0,%
    angley=90,%
    anglez=225,%
    front/.style={draw=red!50!black,fill=red!50!white},%
    top/.style={draw=red!50!black,fill=red!50!white},%
    right/.style={draw=red!50!black,fill=red!50!white},%
    emphedge=false,%
    }
\begin{scope}[shift={(0,0,2)}]
    \tikzcuboid{%
    front/.style={draw=green!50!black,fill=green!50!white},%
    top/.style={draw=green!50!black,fill=green!50!white},%
    right/.style={draw=green!50!black,fill=green!50!white},%
    };   
\end{scope}
\begin{scope}[shift={(0,0,4)}]
    \tikzcuboid{%
    front/.style={draw=blue!50!black,fill=blue!50!white},%
    top/.style={draw=blue!50!black,fill=blue!50!white},%
    right/.style={draw=blue!50!black,fill=blue!50!white},%
    }   
\end{scope}
\begin{scope}[shift={(0,0,6)}]
    \tikzcuboid{%
    front/.style={draw=yellow!50!black,fill=yellow!50!white},%
    top/.style={draw=yellow!50!black,fill=yellow!50!white},%
    right/.style={draw=yellow!50!black,fill=yellow!50!white},%
    }   
\end{scope}

\begin{scope}[shift={(1.5,0,0)}]
    \tikzcuboid{%
    front/.style={draw=lime!50!black,fill=lime!50!white},%
    top/.style={draw=lime!50!black,fill=lime!50!white},%
    right/.style={draw=lime!50!black,fill=lime!50!white},%
    };   
\end{scope}
\begin{scope}[shift={(1.5,0,2)}]
    \tikzcuboid{%
    front/.style={draw=magenta!50!black,fill=magenta!50!white},%
    top/.style={draw=magenta!50!black,fill=magenta!50!white},%
    right/.style={draw=magenta!50!black,fill=magenta!50!white},%
    };   
\end{scope}
\begin{scope}[shift={(1.5,0,4)}]
    \tikzcuboid{%
    front/.style={draw=olive!50!black,fill=olive!50!white},%
    top/.style={draw=olive!50!black,fill=olive!50!white},%
    right/.style={draw=olive!50!black,fill=olive!50!white},%
    }   
\end{scope}
\begin{scope}[shift={(1.5,0,6)}]
    \tikzcuboid{%
    front/.style={draw=pink!50!black,fill=pink!50!white},%
    top/.style={draw=pink!50!black,fill=pink!50!white},%
    right/.style={draw=pink!50!black,fill=pink!50!white},%
    }  % #brown%teal%violet%lime%pink%purple
\end{scope}

\begin{scope}[shift={(3,0,0)}]
    \tikzcuboid{%
    front/.style={draw=white!50!black,fill=white!50!white},%
    top/.style={draw=white!50!black,fill=white!50!white},%
    right/.style={draw=white!50!black,fill=white!50!white},%
    };   
\end{scope}
\begin{scope}[shift={(3,0,2)}]
    \tikzcuboid{%
    front/.style={draw=orange!50!black,fill=orange!50!white},%
    top/.style={draw=orange!50!black,fill=orange!50!white},%
    right/.style={draw=orange!50!black,fill=orange!50!white},%
    };    %purple
\end{scope}
\begin{scope}[shift={(3,0,4)}]
    \tikzcuboid{%
    front/.style={draw=gray!50!black,fill=gray!50!white},%
    top/.style={draw=gray!50!black,fill=gray!50!white},%
    right/.style={draw=gray!50!black,fill=gray!50!white},%
    }   
\end{scope}
\begin{scope}[shift={(3,0,6)}]
    \tikzcuboid{%
    front/.style={draw=cyan!50!black,fill=cyan!50!white},%
    top/.style={draw=cyan!50!black,fill=cyan!50!white},%
    right/.style={draw=cyan!50!black,fill=cyan!50!white},%
    }   
\end{scope}
\end{tikzpicture}
}
\caption{2D Decomposition} \label{fig:m_e}
\end{subfigure}
\begin{subfigure}{0.31\textwidth}
\scalebox{0.5}{
\begin{tikzpicture}
    \tikzcuboid{%
    shiftx=0cm,%
    shifty=0cm,%
    scale=1,%
    rotation=0,%
    densityx=2,%
    densityy=2,%
    densityz=2,%
    dimx=5,%
    dimy=1.5,%
    dimz=1,%
    scalex=1,%
    scaley=1,%
    scalez=0.75,%
    anglex=0,%
    angley=90,%
    anglez=225,%
    front/.style={draw=red!50!black,fill=red!50!white},%
    top/.style={draw=red!50!black,fill=red!50!white},%
    right/.style={draw=red!50!black,fill=red!50!white},%
    emphedge=false,%
    }
\begin{scope}[shift={(0,0,2)}]
    \tikzcuboid{%
    front/.style={draw=green!50!black,fill=green!50!white},%
    top/.style={draw=green!50!black,fill=green!50!white},%
    right/.style={draw=green!50!black,fill=green!50!white},%
    };   
\end{scope}
\begin{scope}[shift={(0,0,4)}]
    \tikzcuboid{%
    front/.style={draw=blue!50!black,fill=blue!50!white},%
    top/.style={draw=blue!50!black,fill=blue!50!white},%
    right/.style={draw=blue!50!black,fill=blue!50!white},%
    }   
\end{scope}
\begin{scope}[shift={(0,0,6)}]
    \tikzcuboid{%
    front/.style={draw=yellow!50!black,fill=yellow!50!white},%
    top/.style={draw=yellow!50!black,fill=yellow!50!white},%
    right/.style={draw=yellow!50!black,fill=yellow!50!white},%
    }   
\end{scope}

\begin{scope}[shift={(0,2,0)}]
    \tikzcuboid{%
    front/.style={draw=lime!50!black,fill=lime!50!white},%
    top/.style={draw=lime!50!black,fill=lime!50!white},%
    right/.style={draw=lime!50!black,fill=lime!50!white},%
    };   
\end{scope}
\begin{scope}[shift={(0,2,2)}]
    \tikzcuboid{%
    front/.style={draw=magenta!50!black,fill=magenta!50!white},%
    top/.style={draw=magenta!50!black,fill=magenta!50!white},%
    right/.style={draw=magenta!50!black,fill=magenta!50!white},%
    };   
\end{scope}
\begin{scope}[shift={(0,2,4)}]
    \tikzcuboid{%
    front/.style={draw=olive!50!black,fill=olive!50!white},%
    top/.style={draw=olive!50!black,fill=olive!50!white},%
    right/.style={draw=olive!50!black,fill=olive!50!white},%
    }   
\end{scope}
\begin{scope}[shift={(0,2,6)}]
    \tikzcuboid{%
    front/.style={draw=pink!50!black,fill=pink!50!white},%
    top/.style={draw=pink!50!black,fill=pink!50!white},%
    right/.style={draw=pink!50!black,fill=pink!50!white},%
    }  % #brown%teal%violet%lime%pink%purple
\end{scope}

\begin{scope}[shift={(0,4,0)}]
    \tikzcuboid{%
    front/.style={draw=white!50!black,fill=white!50!white},%
    top/.style={draw=white!50!black,fill=white!50!white},%
    right/.style={draw=white!50!black,fill=white!50!white},%
    };   
\end{scope}
\begin{scope}[shift={(0,4,2)}]
    \tikzcuboid{%
    front/.style={draw=orange!50!black,fill=orange!50!white},%
    top/.style={draw=orange!50!black,fill=orange!50!white},%
    right/.style={draw=orange!50!black,fill=orange!50!white},%
    };    %purple
\end{scope}
\begin{scope}[shift={(0,4,4)}]
    \tikzcuboid{%
    front/.style={draw=gray!50!black,fill=gray!50!white},%
    top/.style={draw=gray!50!black,fill=gray!50!white},%
    right/.style={draw=gray!50!black,fill=gray!50!white},%
    }   
\end{scope}
\begin{scope}[shift={(0,4,6)}]
    \tikzcuboid{%
    front/.style={draw=cyan!50!black,fill=cyan!50!white},%
    top/.style={draw=cyan!50!black,fill=cyan!50!white},%
    right/.style={draw=cyan!50!black,fill=cyan!50!white},%
    }   
\end{scope}
\end{tikzpicture}
}
\caption{2D Decomposition} \label{fig:m_f}
\end{subfigure}
  \caption{1D/2D/3D decomposition strategies with all their possible states.}
  \label{fig:decomp}
\end{figure}
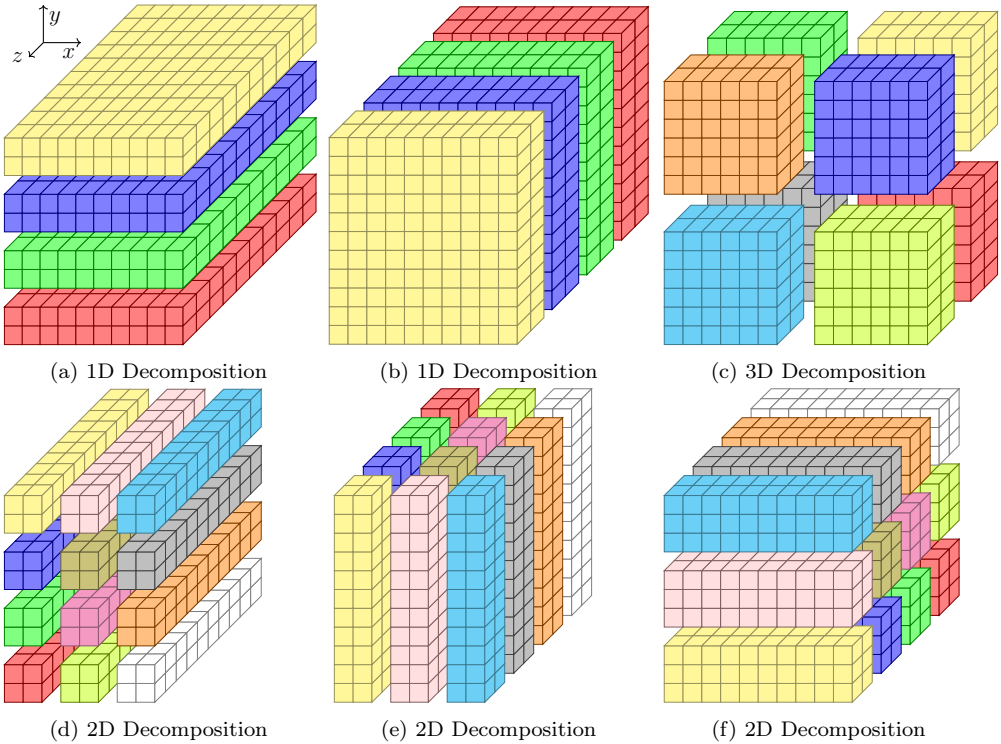

The changes in the non-zero structure in the coefficient matrix $\mathbf{A}$ as a result of forward and backward passes are shown in Figure \ref{fig:thom}.
Additionally, physical systems with periodic boundary conditions lead to a cyclic tridiagonal systems, however, a simple variation of the Thomas algorithm (periodic Thomas) based on the Sherman-Morrison formula can be used to solve these systems \cite{ShermanMorrison}.
Periodic Thomas algorithm implementations require an additional forward pass operation in addition to the forward and backward passes in the Thomas algorithm to obtain the solution.
All these forward and backward passes have a dependency on the previous iteration, thus the Thomas algorithm is a sequential algorithm.
Therefore, it requires all the entries along an individual tridiagonal system to be present in a shared-memory environment such as a single rank.
However, there are various studies where the Thomas algorithm is used for solving batches of tridiagonal systems in 3D domains on multiple ranks \cite{Incompact3D}.
Typically, a 1D or 2D domain decomposition as shown in Figure \ref{fig:decomp} is used in these strategies to distribute batches of tridiagonal systems across multiple ranks.
Additionally, 1D and 2D decomposition strategies alternate between 2 and 3 different states with transposed layouts to have the entirety of the domain in a particular direction in a single rank so that the sequential Thomas algorithm can be applied. 
These transposition operations between decomposition states are also found in multi-rank fast Fourier Transform algorithms \cite{Rolfo2023}.
1D and 2D decompositions and all the states the decomposition alternates in between are demonstrated in Figure \ref{fig:decomp}.
This strategy necessitates MPI all-to-all type communications to carry out transposes across ranks between different states and may cause bottlenecks especially on modern GPU clusters.

\subsection{Distributed Algorithms}
\begin{figure}[h!]
  \centering
\begin{tikzpicture}[scale=0.28]
%\draw (0,0) grid (15,15);
\draw[step=1.0, lightgray] (0,0) grid (15, 15);

\draw[step=5.0, gray] (0,0) grid (15, 15);
%\draw[very thick, step=5.0] (0.1,0.1) grid (14.9, 14.9);
\draw[xshift=10, lightgray] (15,0) grid [xstep=1,ystep=5] (16, 15);
\draw[xshift=10, lightgray] (17,0) grid [xstep=1,ystep=5] (18, 15);
\draw[black, very thick] (-0.4, 5) -- (16.7, 5);
\draw[black, very thick] (-0.4, 10) -- (16.7, 10);
\draw[black, very thick] (17.1, 5) -- (18.7, 5);
\draw[black, very thick] (17.1, 10) -- (18.7, 10);
\foreach \i in {1, 2, ..., 15}{
\filldraw[darkgray] (\i-0.5,15.5-\i) circle (8pt);
}
\foreach \i in {1, 2, ..., 14}{
\filldraw[darkgray] (\i-0.5,14.5-\i) circle (8pt);
\filldraw[darkgray] (\i+0.5,15.5-\i) circle (8pt);
}
\foreach \i in {1, 2, ..., 15}{
\filldraw[xshift=10, blue] (15.5,15.5-\i) circle (8pt);
\filldraw[xshift=10, darkgray] (17.5,15.5-\i) circle (8pt);
}
\node[scale=0.75] at (16.9,7.5) {$=$};

%ranks
\path[draw, very thick, decorate,decoration={calligraphic brace}] (-0.4, 0) -- (-0.4, 5)
node[midway,left]{$\#2$};
\path[draw, very thick, decorate,decoration={calligraphic brace}] (-0.4, 5) -- (-0.4, 10)
node[midway,left]{$\#1$};
\path[draw, very thick, decorate,decoration={calligraphic brace}] (-0.4, 10) -- (-0.4, 15)
node[midway,left]{$\#0$};
\end{tikzpicture}
  \caption{Subdomains in a distributed-memory tridiagonal matrix algorithm. A tridiagonal system with 15 points are divided into 3 subdomains numbered from 0 to 2 each with 5 points. Blue color represents the unknowns.}
  \label{fig:distributed}
\end{figure}
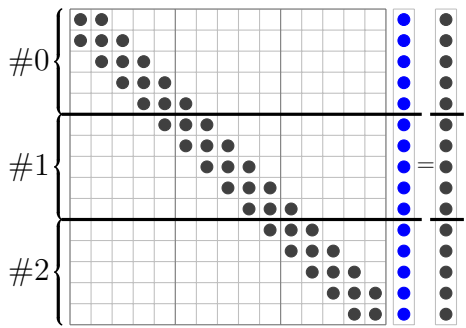

The fundamental difference between a serial algorithm such as the Thomas algorithm and various distributed algorithms is that distributed algorithms divide individual tridiagonal systems into multiple subdomains as shown in Figure \ref{fig:distributed}.
This allows distributed-memory algorithms to have up to 3D domain decomposition as shown in Figure \ref{fig:decomp}, and the decomposition state is constant and does not require frequent alterations between different states as in Thomas algorithm based strategies.
Therefore, the most important advantage of distributed-memory algorithms is that all-to-all type MPI communications involving the entire domain are completely eliminated.

In this subsection, we examine two different distributed algorithms.
The proposed distributed-memory algorithm in this paper for solving tridiagonal systems combines some of the strategies in the two algorithms described in this section and also introduces novel techniques. 

\subsubsection{Hybrid algorithms}
The hybrid Thomas-PCR algorithm was developed by Lazslo et al. in 2016 \cite{Laszlo2016}.
The algorithm consist of 3 phases.
The first phase obtains a reduced sized tridiagonal system where each subdomain contributes two unknowns each.
The primary operations in the first phase of the algorithm are similar to forward and backward passes in the Thomas algorithm, however, they are carried out in the subdomains rather than the entire domain.
Algorithm \ref{alg:lazslo-1} provides the pseudo-code for the first phase of the hybrid algorithm, modified Thomas algorithm, and Figure \ref{fig:laszlo} demonstrates the corresponding changes in the non-zero structure of the coefficient matrix $\mathbf{A}$ as well as highlighting the contributions from each subdomain to the final reduced system.

\begin{singlespacing}
\begin{algorithm}
\caption{Modified Thomas algorithm - first phase.}\label{alg:lazslo-1}
\begin{algorithmic}[1]
\State $u_1 \gets d_1/b_1$
\State $c_1 \gets c_1/b_1$
\For{$i = 2:n$} \Comment{Forward pass}
  \State $w \gets 1/(b_i - a_i c_{i-1})$
  \State $u_i \gets w(d_i - a_i u_{i-1})$  %\Comment{This is a comment}
  \State $c_i \gets wc_i$
\EndFor
\For{$i = n-1:1$} \Comment{Backward pass}
  \State $u_i \gets u_i - c_iu_{i+1}$
\EndFor
\end{algorithmic}
\end{algorithm}
\end{singlespacing}
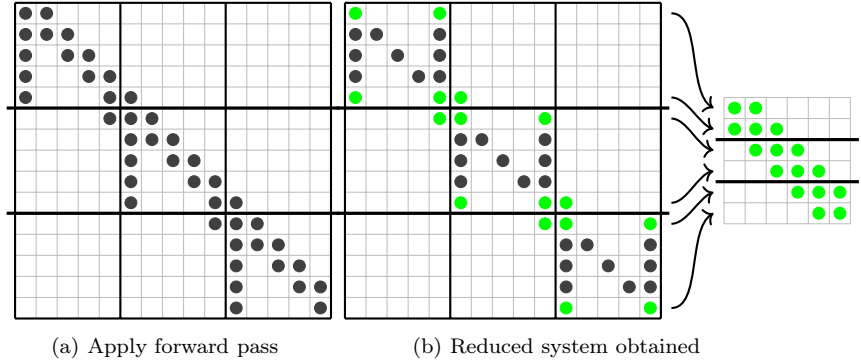
\begin{figure}[h!]
  \centering
\begin{subfigure}{0.31\textwidth}
\begin{tikzpicture}[scale=0.28]
%\draw (0,0) grid (15,15);
\draw[step=1.0, lightgray] (0,0) grid (15, 15);
\draw[step=5.0, thick] (0,0) grid (15, 15);
%\draw[very thick, step=5.0] (0.1,0.1) grid (14.9, 14.9);
%\draw[xshift=10] (15,0) grid [xstep=1,ystep=5] (16, 15);
%\draw[xshift=10] (17,0) grid [xstep=1,ystep=5] (18, 15);
\draw[black, very thick] (-0.4, 5) -- (15.4, 5);%(18.8, 5);
\draw[black, very thick] (-0.4, 10) -- (15.4, 10);
\foreach \i in {1, 2, ..., 15}{
\filldraw[darkgray] (\i-0.5,15.5-\i) circle (8pt);
}
\foreach \i in {1, 2, ..., 14}{
%\filldraw[darkgray] (\i-0.5,14.5-\i) circle (8pt);
\filldraw[darkgray] (\i+0.5,15.5-\i) circle (8pt);
}
\foreach \i in {1, ..., 4}{
\filldraw[darkgray] (0.5,14.5-\i) circle (8pt);
%\filldraw[darkgray] (4.5,15.5-\i) circle (8pt);
}
\foreach \i in {6, ..., 9}{
\filldraw[darkgray] (5.5,14.5-\i) circle (8pt);
%\filldraw[darkgray] (9.5,15.5-\i) circle (8pt);
}
\foreach \i in {11, ..., 14}{
\filldraw[darkgray] (10.5,14.5-\i) circle (8pt);
%\filldraw[darkgray] (14.5,15.5-\i) circle (8pt);
}
\foreach \i in {5, 10}{
\filldraw[darkgray] (\i-0.5,14.5-\i) circle (8pt);
%\filldraw[green] (\i+0.5,15.5-\i) circle (8pt);
}
\end{tikzpicture}
\caption{Apply forward pass} \label{fig:laszlo_a}
\end{subfigure}
\begin{subfigure}{0.43\textwidth}
\begin{tikzpicture}[scale=0.28]
%\draw (0,0) grid (15,15);
\draw[step=1.0, lightgray] (0,0) grid (15, 15);
\draw[step=5.0, thick] (0,0) grid (15, 15);
%\draw[very thick, step=5.0] (0.1,0.1) grid (14.9, 14.9);
%\draw[xshift=10] (15,0) grid [xstep=1,ystep=5] (16, 15);
%\draw[xshift=10] (17,0) grid [xstep=1,ystep=5] (18, 15);
\draw[black, very thick] (-0.4, 5) -- (15.4, 5);%(18.8, 5);
\draw[black, very thick] (-0.4, 10) -- (15.4, 10);
%\foreach \i in {1, 2, ..., 15}{
%\filldraw[darkgray] (\i-0.5,15.5-\i) circle (8pt);
%}
\foreach \i in {2, ..., 4}{
\filldraw[darkgray] (0.5,15.5-\i) circle (8pt);
\filldraw[darkgray] (\i-0.5,15.5-\i) circle (8pt);
\filldraw[darkgray] (4.5,15.5-\i) circle (8pt);
}
\foreach \i in {7, ..., 9}{
\filldraw[darkgray] (5.5,15.5-\i) circle (8pt);
\filldraw[darkgray] (\i-0.5,15.5-\i) circle (8pt);
\filldraw[darkgray] (9.5,15.5-\i) circle (8pt);
}
\foreach \i in {12, ..., 14}{
\filldraw[darkgray] (10.5,15.5-\i) circle (8pt);
\filldraw[darkgray] (\i-0.5,15.5-\i) circle (8pt);
\filldraw[darkgray] (14.5,15.5-\i) circle (8pt);
}
\foreach \i in {5, 10}{
\filldraw[green] (\i-0.5,14.5-\i) circle (8pt);
\filldraw[green] (\i+0.5,15.5-\i) circle (8pt);
}
\foreach \i in {1, 6, 11}{
\filldraw[green] (\i-0.5,15.5-\i) circle (8pt);
\filldraw[green] (\i+3.5,15.5-\i) circle (8pt);
\filldraw[green] (\i-0.5,11.5-\i) circle (8pt);
\filldraw[green] (\i+3.5,11.5-\i) circle (8pt);
}
%\end{tikzpicture}
%\begin{tikzpicture}[scale=0.28]
\def\yup{4.5}
\def\xsh{18}%was 17
\draw[xshift=\xsh cm, yshift=\yup cm, lightgray] (0,0) grid (6,6);
%\draw[step=1.0, lightgray] (0,0) grid (6, 6);
%\draw[step=15.0, thick] (0,0) grid (15, 15);
%\draw[very thick, step=5.0] (0.1,0.1) grid (14.9, 14.9);
%\draw[xshift=10] (15,0) grid [xstep=1,ystep=5] (16, 15);
%\draw[xshift=10] (17,0) grid [xstep=1,ystep=5] (18, 15);
\draw[black, very thick] (-0.4+\xsh, 2+\yup) -- (6.4+\xsh, 2+\yup);
\draw[black, very thick] (-0.4+\xsh, 4+\yup) -- (6.4+\xsh, 4+\yup);
\foreach \i in {1, 2, ..., 6}{
\filldraw[green] (\i-0.5+\xsh,6.5-\i+\yup) circle (8pt);
}
\foreach \i in {1, 2, ..., 5}{
\filldraw[green] (\i-0.5+\xsh,5.5-\i+\yup) circle (8pt);
\filldraw[green] (\i+0.5+\xsh,6.5-\i+\yup) circle (8pt);
}
\draw[->, thick, looseness=0.75] (15.5, 0.5) to [out=0,in=-150] (\xsh-0.5, 5);
\draw[->, thick, looseness=0.75] (15.5, 4.5) to [out=0,in=-160] (\xsh-0.5, 6);
\draw[->, thick, looseness=0.75] (15.5, 5.5) to [out=0,in=-170] (\xsh-0.5, 7);
\draw[->, thick, looseness=0.75] (15.5, 9.5) to [out=0,in=170] (\xsh-0.5, 8);
\draw[->, thick, looseness=0.75] (15.5, 10.5) to [out=0,in=160] (\xsh-0.5, 9);
\draw[->, thick, looseness=0.75] (15.5, 14.5) to [out=0,in=150] (\xsh-0.5, 10);
\end{tikzpicture}
\caption{Reduced system obtained} \label{fig:laszlo_b}
\end{subfigure}
  \caption{First phase of the modified Thomas algorithm. Each subdomain carries out forward and backward passes locally resulting in a reduced system shown with \textcolor{green}{\CIRCLE}.}\label{fig:laszlo} %\bullet
\end{figure}
After the first phases are carried out in each subdomain, a reduced system is obtained where the first and last entries from each subdomain is coupled, forming a smaller tridiagonal system as shown in Figure \ref{fig:laszlo}.
Then, this reduced sized system is solved by using a parallel cyclic reduction (PCR) based strategy \cite{PCR}.
Because the implementation considers a shared memory environment where cores can access data in any subdomain, the PCR based strategy to solve the reduced sized system does not require MPI communications, and the solution can be obtained efficiently.
The final phase after solving the reduced sized system is trivial, and only requires substituting the first and last entries in each subdomain as shown in Algorithm \ref{alg:lazslo-3}.

\begin{singlespacing}
\begin{algorithm}
\caption{Modified Thomas algorithm - substitution phase.}\label{alg:lazslo-3}
\begin{algorithmic}[1]
\For{$i = 2:n-1$} %\Comment{Substitute}
  \State $u_i \gets u_i - a_i u_1 - c_i u_n$
\EndFor
\end{algorithmic}
\end{algorithm}
\end{singlespacing}

Relying on a shared memory environment to solve the reduced sized system effectively limits the algorithm to a single rank, however, extension of the algorithm for the distributed-memory systems was also studied by Balogh et al. \cite{Balogh2022}. 
In the distributed-memory versions, the systems are divided into subdomains and these subdomains are located in different ranks.
In this strategy, solution of the reduced systems can be more costly because they require multiple steps of communication depending on the number of ranks involved.
A different strategy was suggested in %PaScaL\_TDMA algorithm 
\cite{Kim2021}, where the reduced systems are distributed across the participating ranks using all-to-all type communications which significantly reduces the amount of data sent and received. 
In this strategy, every rank solves a batch of reduced systems in its local memory, with no need for a PCR type multiple step communication across ranks.
However the solutions of the reduced systems required to be communicated with all-to-all type communications.

\subsubsection{Parallel diagonally dominant algorithm}
The parallel diagonally dominant algorithm (PDD) was first developed by Sun in 1995 \cite{Sun1995}.
As the name implies, this algorithm can only be used if the tridiagonal system is diagonally dominant.
When a diagonally dominant tridiagonal system is divided into multiple subdomains, the global coupling across these regions can be broken down, and this depends on the strength of the diagonal dominance in the system as well as the sizes of the subdomains as investigated in \cite{Sun1995} in detail.
The algorithm requires inverting the local region of the tridiagonal system that belongs to a rank and multiplying the system and the corresponding section of the RHS array by the local inverse matrix.
The specific steps are provided in Algorithm \ref{alg:PDD}.

\begin{singlespacing}
\begin{algorithm}[!h]
\caption{Parallel diagonally dominant algorithm.}\label{alg:PDD}
\begin{algorithmic}[1]
\State $\mathbf{A}^{inv} \gets \text{invert}(\mathbf{A}[1:n,1:n])$
%$\mathrm{c} = matmul(A^{-1}[:, n], [0$
%\For{$i = 1:n$}
\For{$i = 1:n$}
\State $a_i \gets \mathbf{A}^{inv}[i,n]\cdot c_n$
\State $c_i \gets \mathbf{A}^{inv}[i,1]\cdot c_1$
\EndFor
\end{algorithmic}
\end{algorithm}
\end{singlespacing}

\begin{figure}[h!]
  \centering
\begin{subfigure}{0.31\textwidth}
\begin{tikzpicture}[scale=0.28]
%\draw (0,0) grid (15,15);
\draw[step=1.0, lightgray] (0,0) grid (15, 15);
\draw[step=5.0, thick] (0,0) grid (15, 15);
%\draw[very thick, step=5.0] (0.1,0.1) grid (14.9, 14.9);
%\draw[xshift=10] (15,0) grid [xstep=1,ystep=5] (16, 15);
%\draw[xshift=10] (17,0) grid [xstep=1,ystep=5] (18, 15);
\draw[black, very thick] (-0.4, 5) -- (15.4, 5);
\draw[black, very thick] (-0.4, 10) -- (15.4, 10);
\foreach \i in {1, 2, ..., 15}{
\filldraw[darkgray] (\i-0.5,15.5-\i) circle (8pt);
}
\foreach \i in {1, 2, ..., 14}{
\filldraw[darkgray] (\i-0.5,14.5-\i) circle (8pt);
\filldraw[darkgray] (\i+0.5,15.5-\i) circle (8pt);
}
\end{tikzpicture}
\caption{Initial TDM} \label{fig:pdd_a}
\end{subfigure}
\begin{subfigure}{0.45\textwidth}
\begin{tikzpicture}[scale=0.28]
%\draw (0,0) grid (15,15);
\draw[step=1.0, lightgray] (0,0) grid (15, 15);
\draw[step=5.0, thick] (0,0) grid (15, 15);
%\draw[very thick, step=5.0] (0.1,0.1) grid (14.9, 14.9);
%\draw[xshift=10] (15,0) grid [xstep=1,ystep=5] (16, 15);
%\draw[xshift=10] (17,0) grid [xstep=1,ystep=5] (18, 15);
\draw[black, very thick] (-0.4, 5) -- (15.4, 5);%(18.8, 5);
\draw[black, very thick] (-0.4, 10) -- (15.4, 10);
%\foreach \i in {1, 2, ..., 15}{
%\filldraw[darkgray] (\i-0.5,15.5-\i) circle (8pt);
%}

\foreach \i in {2, ..., 4}{
\filldraw[darkgray] (\i-0.5,15.5-\i) circle (8pt);
\filldraw[darkgray] (5.5,15.5-\i) circle (8pt);
}
\foreach \i in {7, ..., 9}{
\filldraw[darkgray] (4.5,15.5-\i) circle (8pt);
\filldraw[darkgray] (\i-0.5,15.5-\i) circle (8pt);
\filldraw[darkgray] (10.5,15.5-\i) circle (8pt);
}
\foreach \i in {12, ..., 14}{
\filldraw[darkgray] (9.5,15.5-\i) circle (8pt);
\filldraw[darkgray] (\i-0.5,15.5-\i) circle (8pt);
}
\foreach \i in {5, 10}{
\filldraw[green] (\i-0.5,15.5-\i) circle (8pt);
\filldraw[green] (\i-0.5,14.5-\i) circle (8pt);
\filldraw[green] (\i+0.5,15.5-\i) circle (8pt);
\filldraw[green] (\i+0.5,14.5-\i) circle (8pt);
}

\filldraw[green] (0.5,14.5) circle (8pt);
\filldraw[green] (5.5,14.5) circle (8pt);
\filldraw[green] (10.5,9.5) circle (8pt);
\filldraw[green] (4.5,5.5) circle (8pt);
\filldraw[green] (9.5,0.5) circle (8pt);
\filldraw[green] (14.5,0.5) circle (8pt);
%\foreach \i in {1, 6, 11}{
%\filldraw[green] (\i-0.5,15.5-\i) circle (8pt);
%\filldraw[green] (\i+3.5,15.5-\i) circle (8pt);
%\filldraw[green] (\i-0.5,11.5-\i) circle (8pt);
%\filldraw[green] (\i+3.5,11.5-\i) circle (8pt);
%}
%\end{tikzpicture}
%\begin{tikzpicture}[scale=0.28]
\def\yup{4.5}
\def\xsh{18}%was 17
\draw[xshift=\xsh cm, yshift=\yup cm, lightgray] (0,0) grid (6,6);
%\draw[step=1.0, lightgray] (0,0) grid (6, 6);
%\draw[step=15.0, thick] (0,0) grid (15, 15);
%\draw[very thick, step=5.0] (0.1,0.1) grid (14.9, 14.9);
%\draw[xshift=10] (15,0) grid [xstep=1,ystep=5] (16, 15);
%\draw[xshift=10] (17,0) grid [xstep=1,ystep=5] (18, 15);
\draw[black, very thick] (-0.4+\xsh, 2+\yup) -- (6.4+\xsh, 2+\yup);
\draw[black, very thick] (-0.4+\xsh, 4+\yup) -- (6.4+\xsh, 4+\yup);
\foreach \i in {1, 2, ..., 6}{
\filldraw[green] (\i-0.5+\xsh,6.5-\i+\yup) circle (8pt);
}
%\foreach \i in {1, 2, ..., 5}{
%\filldraw[green] (\i-0.5+\xsh,5.5-\i+\yup) circle (8pt);
%\filldraw[green] (\i+0.5+\xsh,6.5-\i+\yup) circle (8pt);
%}
\filldraw[green] (2.5+\xsh,5.5+\yup) circle (8pt);
\filldraw[green] (2.5+\xsh,4.5+\yup) circle (8pt);
\filldraw[green] (1.5+\xsh,3.5+\yup) circle (8pt);
\filldraw[green] (1.5+\xsh,2.5+\yup) circle (8pt);
\filldraw[green] (4.5+\xsh,3.5+\yup) circle (8pt);
\filldraw[green] (4.5+\xsh,2.5+\yup) circle (8pt);
\filldraw[green] (3.5+\xsh,1.5+\yup) circle (8pt);
\filldraw[green] (3.5+\xsh,0.5+\yup) circle (8pt);
\draw[->, thick, looseness=0.75] (15.5, 0.5) to [out=0,in=-150] (\xsh-0.5, 5);
\draw[->, thick, looseness=0.75] (15.5, 4.5) to [out=0,in=-160] (\xsh-0.5, 6);
\draw[->, thick, looseness=0.75] (15.5, 5.5) to [out=0,in=-170] (\xsh-0.5, 7);
\draw[->, thick, looseness=0.75] (15.5, 9.5) to [out=0,in=170] (\xsh-0.5, 8);
\draw[->, thick, looseness=0.75] (15.5, 10.5) to [out=0,in=160] (\xsh-0.5, 9);
\draw[->, thick, looseness=0.75] (15.5, 14.5) to [out=0,in=150] (\xsh-0.5, 10);
\end{tikzpicture}
\caption{TDM multiplied by local inverses} \label{fig:pdd_b}
\end{subfigure}
  \caption{Parallel diagonally dominant algorithm. Each region is multiplied by its local inverse which results in a reduced penta-diagonal system with first and last data points from each rank.}
  \label{fig:pdd}
\end{figure}
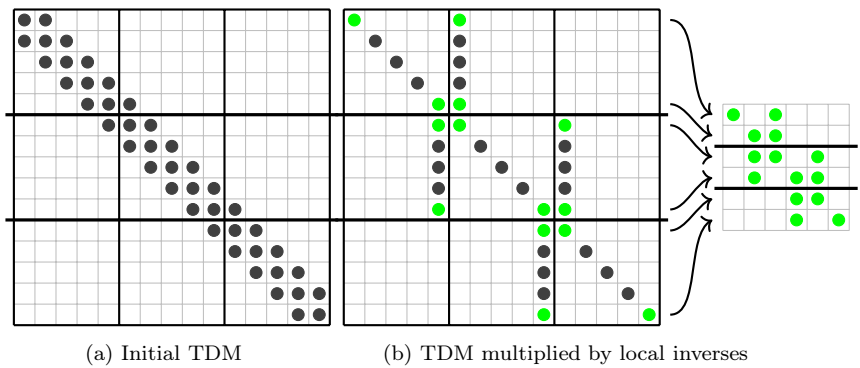

As a result of a multiplication by the inverse, the boundary points in each subdomain decouple from the interior points as shown in Figure \ref{fig:pdd}.
The points in the boundary from each region then form a penta-diagonal system.
However, further analysis in \cite{Sun1995} demonstrates that particular entries in the resulting penta-diagonal system are below the zero-machine value for diagonally dominant systems and can be disregarded without affecting the accuracy of the solution.
Thus, the penta-diagonal systems are transformed into a series of 2x2 systems such that each system is across a subdomain boundary.
As a result, the communication requirements to solve the reduced systems are significantly less compared to an alternative strategy to solve the reduced penta-diagonal systems.

\section{Performance Considerations}\label{sec:perf}
There are two main characteristics in tridiagonal matrix algorithms that are crucial for computational performance. 
First, the FLOP requirements of these algorithms are very low compared to data movement requirements, and therefore they are heavily bandwidth bound.
Thus, achieving high performance requires efficient utilisation of the available bandwidth on a given system.
Second, the algorithms described in Section \ref{sec:TDMA}
include forward and backward passes, where the data points in the domain are accessed and operated on in a sequential way.
The forward and backward passes result in accessing a given data point twice in a short time interval.
Thus, this provides an opportunity to utilise the CPU cache or GPU shared memory to store the intermediate state between forward and backward passes to save unnecessary data movements from and to the main memory.

The frequent accesses to the same memory locations in the algorithm makes it possible to use cache memory and reduce the data movements from the main memory.
However, the sequential nature of the arithmetic operations within the forward and backward passes can disable the vectorisation on CPUs and thread level parallelism on GPUs if not handled correctly.
Although the characteristics at the algorithmic level are promising for exploiting the memory hierarchy and obtaining performance gains, there are a few challenges for achieving these improvements in practice.
First of all, our primary focus is solving tridiagonal systems in 3D domains.
Therefore, both the sizes of the domains and the data structure are very important considerations for enabling locality optimisations. 
For example, a typical data structure for storing a field in a 3D domain is based on a 3D array where $x$, $y$, and $z$ indices enable accessing a location in the domain easily.
However, this results in quite significant jumps in memory between some of the entries that are next to each other in physical space.
Therefore, when we access these entries that are far from each other in memory to perform a numerical operation it may result in indirect memory accesses, potentially limiting the performance.

Exploiting the memory hierarchy via locality optimisations requires the processor to have enough cache memory to store the intermediate results.
Modern CPUs have up to 2 MiB private L2 cache per core, and it is large enough to be used when solving a tridiagonal system in a 3D domain.
Typically, domain sizes along a direction range between 512 and
4096 entries for computational fluid dynamics applications.
An individual tridiagonal system along a given line with 4096 data points in double precision is only 32 KiB in size, and most CPUs can store such systems entirely in cache.
Three new generation CPUs and their specifications are provided in Table \ref{table:CPU-specs} to illustrate current trends.
However, the current generation of GPUs have smaller cache sizes compared to CPUs as shown in Table \ref{table:GPU-specs}. Depending on the size of cache requirement of the operations which need to be carried out, GPU caches may not be large enough, leaving kernel fusion as the only option for reducing data movements on GPUs.
\begin{table}[h!]
\centering
\caption{Specifications of state of the art CPUs in 2024 by AMD, Intel, and Nvidia.}\label{table:CPU-specs}
\scalebox{0.8}{
\begin{tabular}{l|cccc}
 & AMD EPYC 9754 & Intel 8490H & Nvidia Grace CPU\\ \hline
Microarchitecture & Zen 4c & Sapphire Rapids & Neoverse V2 \\
\# of cores            & 128 & 60 & 72 \\
L1d Cache           & 32KiB & 48KiB & 64KiB  \\
L2 Cache (private)        & 1MiB & 2MiB & 1MiB \\
Peak Memory BW     & 460.8 GB/s & 307.2 GB/s & 512GB/s \\
%Peak FLOP/s                 &   
\end{tabular}
}
\end{table}
\begin{table}[h!]
\centering
\caption{Specifications of state of the art GPUs in 2024 by AMD, Intel, and Nvidia.}\label{table:GPU-specs}
\scalebox{0.8}{
\begin{tabular}{l|cccc}
 & AMD MI300X & Intel GPU Max & Nvidia GH200 \\ \hline
Architecture & CDNA3 & GPU Max & Hopper \\ 
\# of SM/CU/XeC            & 304 & 128 & 132 \\
\# of DP cores per SM     & 32 & 32 & 64 \\
Total Memory & 192GB & 128GB & 96GB \\
Total L1 Cache           & 80KiB & 512KiB & 264KiB  \\
Shared L1 (Scratch)      & 64KiB & 512KiB & 228KiB  \\
L2 Cache (shared)        & $8\times 4 $MiB & 288MiB & 50MiB  \\
Peak Memory BW     & 5.3TB/s & 3.2TB/s & 4TB/s \\
%Peak FLOP/s                 &   
\end{tabular}
}
\end{table}

Finally, vectorisation on CPUs and thread level parallelisation on GPUs are also important factors to make efficient use of a given hardware.
For example, sequential operations in the Thomas and Modified Thomas algorithms can prevent vectorisation and thread level parallelisation based on the data layout.
If the data points that need to be operated on sequentially are stored next to each other in memory, CPUs cannot issue vector instructions such as AVX512.
Moreover, threads in a GPU thread block are effectively parallelised and work efficiently only when they can operate concurrently, which does not align well with sequential operations.

To summarise, the data access pattern and exploiting the memory hierarchy are critical for achieving a high performance implementation for tridiagonal matrix solvers.
First, a better data structure can lead to a more linearised data access pattern, enabling a higher percentage of peak bandwidth. 
Also, a well designed data structure can enable vectorisation on CPUs and thread level parallelisation on GPUs for improved performance.

\section{Proposed Strategy}\label{sec:method}
In this section, we provide a detailed description of a specialised data structure along with a new distributed-memory tridiagonal solver algorithm, DistD2-TDS, to achieve high performance on CPU and GPU hardware. 
Furthermore, we also discuss the advantages of the data structure in terms of performance improvements both for a Thomas algorithm implementation and for our custom distributed-memory tridiagonal solver implementation based on DistD2 algorithm.
\subsection{Data Structure}
As mentioned in Section \ref{sec:intro} and further discussed in Section \ref{sec:TDMA}, tridiagonal solver algorithms often require sequential operations.
However, we are interested in solving batches of tridiagonal systems in a 3D domain such that the individual tridiagonal systems are independent from each other.
Although this provides a decent level of parallelism that can be taken advantage of, a naive implementation might still lead to underutilisation of the hardware resulting in low performance due to the sequential operations required by these algorithms. 
In particular, $x$-directional tridiagonal systems are the most vulnerable because of the typical Cartesian data structure where data points along the $x$-direction are stored next to each other in memory, while in $y$- and $z$-directions there is a jump in memory between data points that are adjacent to each other in physical domain.
In case of sequential data dependencies between data points that are stored next to each other in memory, CPUs cannot use vector instructions and GPUs cannot easily take advantage of thread level parallelism.
Thus, $x$-directional tridiagonal systems typically suffer from low performance.
On the other hand, $y$- and $z$-directional tridiagonal systems can use vector instructions on CPUs and also fully utilise the available thread parallelism on GPUs.
However, because the physically adjacent data points are separated by significant jumps in memory, the data access pattern in $y$- and $z$-directional tridiagonal systems is non-contiguous and this may cause a reduced performance.

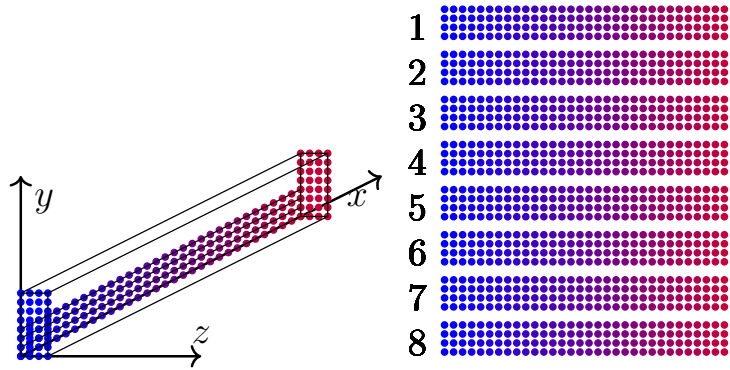
\begin{figure}[h!]
  \centering
\scalebox{1.2}{
\begin{tikzpicture}[x = {(1cm,0.5cm)}, y={(0cm,1cm)}, z={(1cm,0cm)},]

    \draw[thick,->] (0,0,0) -- (4,0,0) node[anchor=north east]{$x$};
    \draw[thick,->] (0,0,0) -- (0,2,0) node[anchor=north west]{$y$};
    \draw[thick,->] (0,0,0) -- (0,0,2) node[anchor=south]{$z$};
\def\dx{0.1}
\def\dy{0.1}
\def\dz{0.1}
\def\nx{31}
\def\ny{7}
\def\nz{3}

%\foreach \x in {0,...,31}
%  \foreach \y in {0,...,7}
%    \foreach \z in {0,...,3}
%      \draw [color=orange, mark=*, mark size=1] plot coordinates{(\x*\dx, \y*\dy, \z*\dz)};

\foreach \x in {0,...,31}{
  \foreach \y in {0,...,3}{
    \foreach \z in {0}{
    \pgfmathsetmacro\k{\x*2.5}
      \draw [color=red!\k!blue, mark=*, mark size=1] plot coordinates{(\x*\dx, \y*\dy, \z*\dz)};}}}
\foreach \x in {0,\nx}{
  \foreach \y in {0,...,7}{
    \foreach \z in {0,...,3}{
    \pgfmathsetmacro\k{\x*2.5}
      \draw [color=red!\k!blue, mark=*, mark size=1] plot coordinates{(\x*\dx, \y*\dy, \z*\dz)};}}}

\path (0,0,0) coordinate (A) (0,0,\nz*\dz) coordinate (B) (0,\ny*\dy,0) coordinate (C) (0,\ny*\dy,\nz*\dz) coordinate (D); 
%(0,\z,\y) coordinate (E) (\x,\z,\y) coordinate (F) (\x,\z,0) coordinate (G) (0,\z,0) coordinate (H);
\path (\nx*\dx,0,0) coordinate (E) (\nx*\dx,0,\nz*\dz) coordinate (F) (\nx*\dx,\ny*\dy,0) coordinate (G) (\nx*\dx,\ny*\dy,\nz*\dz) coordinate (H); 
%\path (-1,-1,1) coordinate (E) (1,-1,1) coordinate (F) (-1,1,1) coordinate (G)  (-1,1,1) coordinate (H);
\draw (A)--(B)--(D)--(C)--(A);%--(D)--(E)--(F)--(D);
\draw (E)--(F)--(H)--(G)--(E);%--(D)--(E)--(F)--(D);
\draw (A)--(E);
\draw (B)--(F);
\draw (C)--(G);
\draw (D)--(H);
\draw (0, \dy, 0) -- (\nx*\dx, \dy, 0);
\draw (0, 2*\dy, 0) -- (\nx*\dx, 2*\dy, 0);
\draw (0, 3*\dy, 0) -- (\nx*\dx, 3*\dy, 0);
\end{tikzpicture}
\begin{tikzpicture}
\def\dx{0.1}
\def\dy{0.1}
\def\dz{0.1}
\def\nx{31}
\def\ny{7}
\def\nz{3}
\definecolor{redw}{rgb}{1,0.45,0.45}
\definecolor{bluew}{rgb}{0.45,0.45,1}
%\foreach \i in {1, 2, ..., 15}{
%\pgfmathsetmacro\k{\i*5}
%\filldraw[xshift=10, redw!\k!bluew] (20+\i,6) circle (8pt);
\foreach \i in {1,...,8}{
\foreach \x in {0,...,31}{
  \node[](Group\i) at (-0.3,1.5*8*4*\dy-0.2-\i*4*\dy-\i*0.1) {\i};
  \foreach \y in {0,...,3}{
    \foreach \z in {0}{
    \pgfmathsetmacro\k{\x*2.5}
      \draw [color=red!\k!blue, mark=*, mark size=1] plot coordinates{(\x*\dx, \y*\dy+\i*4*\dy+\i*0.1, \z*\dz)};}}}}
\end{tikzpicture}
}
  \caption{The proposed data structure for an $x$-directional tridiagonal system. Data continuity in memory is in column-major order.}
  \label{fig:pnclgroup}
\end{figure}

In order to enable a linear and predictable memory access pattern regardless of the spatial direction of the tridiagonal systems, we subdivide the computational domain into groups of individual tridiagonal systems  and tightly pack these individual systems so that their data is contiguous in memory.
A schematic is provided in Figure \ref{fig:pnclgroup} where the domain size is $32 \times 8\times 4$, here a single group consists of 4 individual tridiagonal systems and thus the domain can be mapped in 8 groups as numbered in the figure.
Thus, a Cartesian mesh with $n_x$, $n_y$, and $n_z$ entries is mapped as $(SZ, n_x, n_y\cdot n_z/SZ)$ with the proposed data structure in the $x$-directional data layout.
The optimum number of tridiagonal systems in a group, $SZ$, depends on the specifics of the hardware, and is 4 in Figure \ref{fig:pnclgroup} for the simplicity of demonstration.
Typically, for a double precision simulation on CPUs, $SZ=8$ as vector registers are typically 512 bit long and they can be used to carry out $8$ double precision FLOPs per cycle, and for a double precision simulation on GPUs, $SZ=32$ as a single streaming multiprocessor (SM) typically has 32 double precision cores in total, where each core is effectively assigned an individual tridiagonal system.

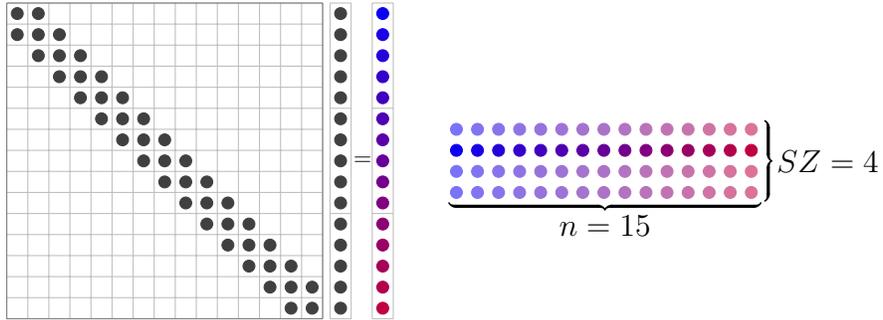
\begin{figure}[h!]
  \centering
\begin{tikzpicture}[scale=0.28]
%\draw (0,0) grid (15,15);
\draw[step=1.0, lightgray] (0,0) grid (15, 15);
\draw[step=15.0, gray] (0,0) grid (15, 15);
%\draw[very thick, step=5.0] (0.1,0.1) grid (14.9, 14.9);
\draw[xshift=10, lightgray] (15,0) grid [xstep=1,ystep=5] (16, 15);
\draw[xshift=10, lightgray] (17,0) grid [xstep=1,ystep=5] (18, 15);
%\draw[black, very thick] (-0.4, 5) -- (16.7, 5);
%\draw[black, very thick] (-0.4, 10) -- (16.7, 10);
%\draw[black, very thick] (17.1, 5) -- (18.7, 5);
%\draw[black, very thick] (17.1, 10) -- (18.7, 10);
\foreach \i in {1, 2, ..., 15}{
\filldraw[darkgray] (\i-0.5,15.5-\i) circle (8pt);
}
\foreach \i in {1, 2, ..., 14}{
\filldraw[darkgray] (\i-0.5,14.5-\i) circle (8pt);
\filldraw[darkgray] (\i+0.5,15.5-\i) circle (8pt);
}
\foreach \i in {1, 2, ..., 15}{
\pgfmathsetmacro\k{\i*5}
\filldraw[xshift=10, darkgray] (15.5,15.5-\i) circle (8pt);
\filldraw[xshift=10, red!\k!blue] (17.5,15.5-\i) circle (8pt);
}
\node[scale=0.75] at (16.9,7.5) {$=$};

\definecolor{redw}{rgb}{1,0.45,0.45}
\definecolor{bluew}{rgb}{0.45,0.45,1}
\foreach \i in {1, 2, ..., 15}{
\pgfmathsetmacro\k{\i*5}
\filldraw[xshift=10, redw!\k!bluew] (20+\i,6) circle (8pt);
\filldraw[xshift=10, redw!\k!bluew] (20+\i,7) circle (8pt);
\filldraw[xshift=10, red!\k!blue] (20+\i,8) circle (8pt);
\filldraw[xshift=10, redw!\k!bluew] (20+\i,9) circle (8pt);
}
\path[draw, very thick, decorate,decoration={calligraphic brace, mirror}] (21,5.5) -- (35.8, 5.5)
node[midway,below]{$n = 15$};
\path[draw, very thick, decorate,decoration={calligraphic brace, mirror}] (36, 5.6) -- (36, 9.4)
node[midway,right]{$SZ = 4$};
%\draw [decorate,
    decoration = {calligraphic brace}] (0.2,1) --  (1.2,1);
\end{tikzpicture}

%\begin{tikzpicture}
%\def\dx{0.1}
%\def\dy{0.1}
%\def\dz{0.1}
%\def\nx{31}
%\def\ny{7}
%\def\nz{3}
%\foreach \x in {0,...,31}{
%  \foreach \y in {0,...,3}{
%    \foreach \z in {0}{
%    \pgfmathsetmacro\k{\x*5}
%      \draw [color=red!\k!gray, mark=*, mark size=1] plot coordinates{(\x*\dx, \y*\dy, \z*\dz)};}}}
%\end{tikzpicture}
  \caption{A single group with $SZ=4$ and $n=15$. The tridiagonal system is demonstrated with a single RHS vector, and the location of the RHS vector in the specialised data structure is highlighted in colour to indicate the exact distribution of a single RHS in memory. The data layout is in column-major order, thus each subsequent data point in a single tridiagonal system is separated from each other by $SZ$ many data points in memory.}
  \label{fig:datastrct}
\end{figure}

In order to enable vectorisation and thread level parallelism the proposed data structure packs the $n^{th}$ entries of all the tridiagonal systems in a single group next to each other in memory.
Therefore, the sequential operations in the algorithm as we apply the forward and backward passes can be concurrently executed for $SZ$ systems at once per core on a CPU and per SM on a GPU.
Furthermore, the substitution phase of the hybrid algorithm also benefits from the data structure in a similar way.
Figure \ref{fig:datastrct} demonstrates how a tridiagonal system is stored in memory with the proposed data structure.
One important note here is that the data layout is specific to a direction in the domain, and it requires reordering based on the direction of the tridiagonal solver.
However, reordering operations account for only a small percentage of the total runtime as demonstrated later in Section \ref{sec:3D-PDE}.

Apart from enabling the vectorisation on CPUs and thread parallelism on GPUs, there are two more main benefits of the proposed data structure.
First, the data structure packs only a small number of lines together, and the tridiagonal solver implementation makes sure that all the lines packed together are being solved concurrently as the data is accessed from main memory.
Thus, the data access is linear and predictable, with no intermittent memory accesses that would be present in $y$- or $z$-directional tridiagonal solvers in a Cartesian data structure.
Second, the tight packing of lines that are concurrently being solved also makes it possible to enable a more efficient cache utilisation based on temporal locality between the forward and backward passes of the tridiagonal solver algorithms.
A similar tight data packing strategy is used in \cite{Akkurt2022} to take advantage of temporal localities and enable cache blocking.

\subsection{DistD2-TDS - Customised tridiagonal matrix solver algorithm}
In this subsection, we describe our customised distributed-memory tridiagonal solver algorithm in detail, and discuss its advantages over existing methodologies.
First of all, we focus on solving tridiagonal systems resulting from high-order compact finite difference schemes \cite{Lele1992}.
A compact scheme for a sixth-order accurate approximation to a first-order derivative can be given as
\begin{equation}\label{eq:cfds}
\alpha u'_{i-1} + u'_i + \alpha u'_{i+1}
=
a\frac{u_{i+1} - u_{i-1}}{2h}
+
b\frac{u_{i+2} - u_{i-2}}{4h},
\end{equation}
where $u$ are the field entries, $u'$ is the derivative, $h$ is the grid spacing, $\alpha$, $a$, and $b$ are the coefficients that are obtained from the Taylor series expansions to satisfy sixth-order accuracy ($\alpha=\nicefrac{1}{3}$, $a=\nicefrac{14}{9}$, $b=\nicefrac{1}{9}$).
The formulation on the right hand side of the equation (RHS) is the $\mathbf{d}$ vector in Equation \ref{eq:TDS}.
Compact schemes can also be used to obtain higher-order derivatives, interpolations and filters \cite{Lele1992}. High-order implicit finite difference schemes offer several advantages in computational fluid dynamics (CFD) for solving partial differential equations (PDEs). 
They can achieve greater accuracy for the same grid resolution compared to lower-order schemes. This is particularly beneficial for capturing fine details and complex flow structures in fluid dynamics simulations. 
They generally exhibit low numerical dissipation and dispersion, preserving the physical fidelity of the simulation. 
Although high-order implicit schemes involve more complex computations per grid point, their ability to achieve high accuracy with fewer grid points can lead to overall computational savings \cite{Lele1992}. 
A good compromise in terms of cost and accuracy are sixth-order implicit schemes which are based on triadiagonal solvers.
These types of schemes are used for instance in Xcompact3d, a framework of finite-difference solvers, designed to study fluid flow problems on supercomputers \cite{Incompact3D}. 
As an example, the solvers in Xcompact3d require approximately 150 tridiagonal systems to be solved at each time step when simulating a fluid flow problem.
One important characteristic is that the tridiagonal systems resulting from high-order compact finite difference schemes are always diagonally dominant, and the proposed algorithm is taking advantage of the diagonal dominance to minimise the communication requirements between ranks.

Before describing the details of our customised strategy, we first investigate the possibility of preprocessing the coefficient arrays to avoid recalculating them on the fly. 
Applying a compact scheme formulation in a 3D domain along a given direction requires solving a tridiagonal system with many RHS arrays.
Therefore, we start by preprocessing the tridiagonal system that defines the operation into a set of coefficient arrays 
such that these preprocessed arrays can be used to avoid recalculating them on the fly for each RHS array.
Algorithm \ref{alg:coeff_prep} describes the procedure, which has to be carried out only once for each distinct operation that correspond to a tridiagonal matrix.

\begin{singlespacing}
\begin{algorithm}[h!]
\caption{Preprocessing for the DistD2 algorithm.}\label{alg:coeff_prep}
\begin{algorithmic}[1]
\For{$j = 1:2$}
\State $s^a_j = s^a_j/b_j$
\State $s^c_j = s^c_j/b_j$
\State $w_j = s^c_j$
\State $f_j = 1/b_j$
\EndFor
\For{$j = 3:n$}
\State $w_j = 1/(b_j - s^a_j s^c_{j-1})$
\State $r_j = s^a_j$
\State $s^a_j = -w_j s^a_j s^a_{j-1}$
\State $s^c_j = w_j s^c_j$
\EndFor
\For{$j = n - 2:2:-1$}
\State $s^a_j = s^a_j - s^c_j s^a_{j+1}$
\State $w_j = s^c_j$
\State $s^c_j = - s^c_j s^c_{j+1}$
\EndFor
\State $w_1 = 1/(1 - s^c_1 s^a_2$
\State $s^a_1 = w_1 s^a_1$
\State $s^c_1 = -w_1 s^c_1 s^c_2$
\end{algorithmic}
\end{algorithm}
\end{singlespacing}

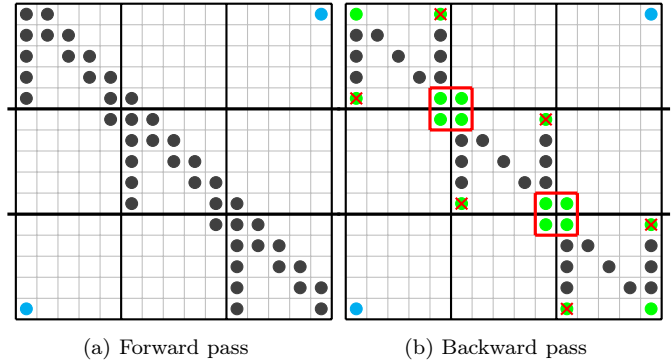
\begin{figure}[h!]
  \centering
\begin{subfigure}{0.31\textwidth}
\begin{tikzpicture}[scale=0.28]
%\draw (0,0) grid (15,15);
\draw[step=1.0, lightgray] (0,0) grid (15, 15);
\draw[step=5.0, thick] (0,0) grid (15, 15);
%\draw[very thick, step=5.0] (0.1,0.1) grid (14.9, 14.9);
%\draw[xshift=10] (15,0) grid [xstep=1,ystep=5] (16, 15);
%\draw[xshift=10] (17,0) grid [xstep=1,ystep=5] (18, 15);
\draw[black, very thick] (-0.4, 5) -- (15.4, 5);%(18.8, 5);
\draw[black, very thick] (-0.4, 10) -- (15.4, 10);
\foreach \i in {1, 2, ..., 15}{
\filldraw[darkgray] (\i-0.5,15.5-\i) circle (8pt);
}
\foreach \i in {1, 2, ..., 14}{
%\filldraw[darkgray] (\i-0.5,14.5-\i) circle (8pt);
\filldraw[darkgray] (\i+0.5,15.5-\i) circle (8pt);
}
\foreach \i in {1, ..., 4}{
\filldraw[darkgray] (0.5,14.5-\i) circle (8pt);
%\filldraw[darkgray] (4.5,15.5-\i) circle (8pt);
}
\foreach \i in {6, ..., 9}{
\filldraw[darkgray] (5.5,14.5-\i) circle (8pt);
%\filldraw[darkgray] (9.5,15.5-\i) circle (8pt);
}
\foreach \i in {11, ..., 14}{
\filldraw[darkgray] (10.5,14.5-\i) circle (8pt);
%\filldraw[darkgray] (14.5,15.5-\i) circle (8pt);
}
\foreach \i in {5, 10}{
\filldraw[darkgray] (\i-0.5,14.5-\i) circle (8pt);
%\filldraw[green] (\i+0.5,15.5-\i) circle (8pt);
}
\filldraw[cyan] (14.5,14.5) circle (8pt);
\filldraw[cyan] (0.5,0.5) circle (8pt);
\end{tikzpicture}
\caption{Forward pass} \label{fig:alg_a}
\end{subfigure}
\begin{subfigure}{0.31\textwidth}
\begin{tikzpicture}[scale=0.28]
%\draw (0,0) grid (15,15);
\draw[step=1.0, lightgray] (0,0) grid (15, 15);
\draw[step=5.0, thick] (0,0) grid (15, 15);
%\draw[very thick, step=5.0] (0.1,0.1) grid (14.9, 14.9);
%\draw[xshift=10] (15,0) grid [xstep=1,ystep=5] (16, 15);
%\draw[xshift=10] (17,0) grid [xstep=1,ystep=5] (18, 15);
\draw[black, very thick] (-0.4, 5) -- (15.4, 5);%(18.8, 5);
\draw[black, very thick] (-0.4, 10) -- (15.4, 10);
%\foreach \i in {1, 2, ..., 15}{
%\filldraw[darkgray] (\i-0.5,15.5-\i) circle (8pt);
%}
\foreach \i in {2, ..., 4}{
\filldraw[darkgray] (0.5,15.5-\i) circle (8pt);
\filldraw[darkgray] (\i-0.5,15.5-\i) circle (8pt);
\filldraw[darkgray] (4.5,15.5-\i) circle (8pt);
}
\foreach \i in {7, ..., 9}{
\filldraw[darkgray] (5.5,15.5-\i) circle (8pt);
\filldraw[darkgray] (\i-0.5,15.5-\i) circle (8pt);
\filldraw[darkgray] (9.5,15.5-\i) circle (8pt);
}
\foreach \i in {12, ..., 14}{
\filldraw[darkgray] (10.5,15.5-\i) circle (8pt);
\filldraw[darkgray] (\i-0.5,15.5-\i) circle (8pt);
\filldraw[darkgray] (14.5,15.5-\i) circle (8pt);
}
\foreach \i in {5, 10}{
\filldraw[green] (\i-0.5,14.5-\i) circle (8pt);
\filldraw[green] (\i+0.5,15.5-\i) circle (8pt);
}
\foreach \i in {1, 6, 11}{
\filldraw[green] (\i-0.5,15.5-\i) circle (8pt);
%\filldraw[green, cross out, draw=red] (\i+3.5,15.5-\i) circle (8pt);
\filldraw[green] (\i+3.5,15.5-\i) circle (8pt);
\draw (\i+3.5,15.5-\i) node[cross=3pt,red, thick] {};
%\draw[cross out, draw=red] circle (8pt);
%\node[cross out, draw] at (\i+3.5,15.5-\i) {o};
\filldraw[green] (\i-0.5,11.5-\i) circle (8pt);
\draw (\i-0.5,11.5-\i) node[cross=3pt,red, thick] {};
\filldraw[green] (\i+3.5,11.5-\i) circle (8pt);
}

\draw[step=2.0, xshift=1cm, very thick, red] (8,4) grid (10, 6);
\draw[step=2.0, yshift=1cm, very thick, red] (4,8) grid (6, 10);

\filldraw[cyan] (14.5,14.5) circle (8pt);
\filldraw[cyan] (0.5,0.5) circle (8pt);
\end{tikzpicture}
\caption{Backward pass} \label{fig:alg_b}
\end{subfigure}
  \caption{Distributed-memory tridiagonal matrix algorithm. Forward and backward passes as in modified Thomas algorithm result in reduced system shown with \textcolor{green}{\CIRCLE}. \textcolor{cyan}{\CIRCLE} indicate the non-zero entries in a cyclic tridiagonal system, and if they are present they remain unchanged in forward and backward passes. A diagonally dominant system results in below zero-machine values in certain locations, marked with \textcolor{red}{$\boldsymbol{\times}$} to indicate that they are disregarded. Red squares show the decoupled 2$\times$2 systems across the domain. In case of a cyclic system, the non-zero entries at the corners form an additional 2$\times$2 system between the first and last ranks.}
  \label{fig:alg}
\end{figure}

\begin{figure}[h!]
  \centering
\begin{tikzpicture}[scale=0.28]
\definecolor{redw}{rgb}{1,0.45,0.45}
\definecolor{bluew}{rgb}{0.45,0.45,1}
\foreach \i in {1, 2, ..., 15}{
\pgfmathsetmacro\k{\i*5}
\filldraw[red!\k!blue] (\i,14) circle (8pt);
\filldraw[red!\k!blue] (\i,13) circle (8pt);
\filldraw[red!\k!blue] (\i,12) circle (8pt);
\filldraw[red!\k!blue] (\i,11) circle (8pt);
}
\path[draw, very thick, decorate,decoration=calligraphic brace] (0.5,10.5) -- (0.5, 14.5)
node[above=-16pt, rotate=90, align=center]{\tiny Input field};

%\path[draw, very thick, decorate,decoration={calligraphic brace, mirror}] (21,5.5) -- (35.8, 5.5) 
%node[midway,below]{$n = 15$};
%\path[draw, very thick, decorate,decoration={calligraphic brace, mirror}] (36, 5.6) -- (36, 9.4) 
%node[midway,right]{$SZ = 4$};
%\draw [decorate, decoration = {calligraphic brace}] (0.2,1) --  (1.2,1);

\path[draw, very thick, decorate,decoration=calligraphic brace] (5.5,15) -- (10.5, 15)
node[midway,above]{\tiny $(j-2 : j+2)$};

\draw[draw=black, dashed] (5.5,10.5) rectangle ++(5,4);
\draw[black, dashed] (5.5, 10.5) -- (7.5, 9.5);
\draw[black, dashed] (10.5, 10.5) -- (8.5, 9.5);

\foreach \i in {8}{
\pgfmathsetmacro\k{\i*5}
\filldraw[red!\k!blue] (\i,9) circle (8pt);
\filldraw[red!\k!blue] (\i,8) circle (8pt);
\filldraw[red!\k!blue] (\i,7) circle (8pt);
\filldraw[red!\k!blue] (\i,6) circle (8pt);
}
%\node[draw,dotted,fit=(8,6) (8,9) {};

\node [rotate=90, align=center] at (6.5,7.5) (noderhs) {\tiny RHS};
\draw[draw=black, dashed] (7.5,5.5) rectangle ++(1,4);

\draw[->] (10.5,7.5) -- (15,7.5) node[midway,below]{\scriptsize Forward pass};

\foreach \i in {1, 2, ..., 8}{
\pgfmathsetmacro\k{\i*5}
\filldraw[red!\k!blue] (\i,4) circle (8pt);
\filldraw[red!\k!blue] (\i,3) circle (8pt);
\filldraw[red!\k!blue] (\i,2) circle (8pt);
\filldraw[red!\k!blue] (\i,1) circle (8pt);
}
\foreach \i in {9, 10, ..., 15}{
\pgfmathsetmacro\k{\i*5}
\draw[red!\k!blue] (\i,4) circle (8pt);
\draw[red!\k!blue] (\i,3) circle (8pt);
\draw[red!\k!blue] (\i,2) circle (8pt);
\draw[red!\k!blue] (\i,1) circle (8pt);
}
\path[draw, very thick, decorate,decoration=calligraphic brace] (0.5,0.5) -- (0.5, 4.5)
node[above=-16pt, rotate=90, align=center, font=\tiny]{Intermediate\\ state};

\draw[draw=black, dashed, very thick] (6.5,0.5) rectangle ++(1,4);
\draw[draw=black] (7.5,0.5) rectangle ++(1,4);

\draw[black, dotted] (8, 0) -- (8, 15);

\node at (8, -1) (nodej) {\scriptsize $j=8$};

\end{tikzpicture}

%\begin{tikzpicture}
%\def\dx{0.1}
%\def\dy{0.1}
%\def\dz{0.1}
%\def\nx{31}
%\def\ny{7}
%\def\nz{3}
%\foreach \x in {0,...,31}{
%  \foreach \y in {0,...,3}{
%    \foreach \z in {0}{
%    \pgfmathsetmacro\k{\x*5}
%      \draw [color=red!\k!gray, mark=*, mark size=1] plot coordinates{(\x*\dx, \y*\dy, \z*\dz)};}}}
%\end{tikzpicture}
  \caption{Visualisation of the fused forward pass and RHS construction operation. Input field is shown on top, and the intermediate state is at the bottom. The RHS at each $j$ is constructed using the input field values at $j-2$ to $j+2$. The forward pass is then applied using the intermediate state values at $j-1$. $SZ$ lines are processed concurrently in a CPU core or in a GPU SM.}
  \label{fig:traverse}
\end{figure}
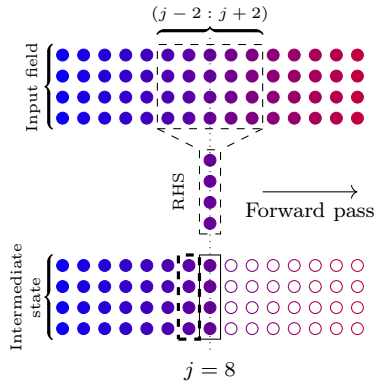

The proposed strategy starts by dividing a batch of tridiagonal systems into multiple subdomains as in Figure \ref{fig:distributed}, where the subdomains are located across multiple ranks in a distributed memory environment.
There are two novel aspects in DistD2 algorithm.
First, it uses a specialised data structure 
to enable vectorisation on CPUs and thread level parallelism on GPUs.
Second, we fuse the RHS construction based on Equation \ref{eq:cfds} with the forward pass in the decoupling phase of the algorithm such that the data movement requirements are minimised.
Using the preprocessed coefficient arrays described in Algorithm \ref{alg:coeff_prep}, decoupling phase of DistD2 algorithm based on the specialised data structure can be described as shown in Algorithm \ref{alg:CTDS}, and the corresponding changes in the non-zero structure of the tridiagonal matrix are shown in Figure \ref{fig:alg}.
Additionally, Figure \ref{fig:traverse} illustrates the fused RHS construction and forward pass, and also emphasises the concurrent execution of $SZ$ many systems at once per core on a CPU or per SM on a GPU.

\begin{singlespacing}
\begin{algorithm}[h!]
\caption{DistD2 - decoupling phase.}\label{alg:CTDS}
\begin{algorithmic}[1]
\For{$j = 1:2$}
\For{$i = 1:SZ$}\Comment{Vectorised Loop}
\State $d_{i,j} = c_{-2,j} u_{i,j-2} + c_{-1,j} u_{i,j-1} + c_{0,j} u_{i,j} + c_{1,j} u_{i,j+1} + c_{2,j} u_{i,j+2}$
\State $d_{i,j} = d_{i,j} r_j$
\EndFor
\EndFor
\For{$j = 3:n$}
\For{$i = 1:SZ$}\Comment{Vectorised Loop}
\State $d_{i,j} = c_{-2,j} u_{i,j-2} + c_{-1,j} u_{i,j-1} + c_{0,j} u_{i,j} + c_{1,j} u_{i,j+1} + c_{2,j} u_{i,j+2}$
\State $d_{i,j} = f_j(d_{i,j} - r_i d_{i,j-1})$
\EndFor
\EndFor
\For{$j = n - 2:2:-1$}
\For{$i = 1:SZ$}\Comment{Vectorised Loop}
\State $d_{i,j} = d_{i,j} - w_j d_{i,j+1}$
\EndFor
\EndFor
\For{$i = 1:SZ$}\Comment{Vectorised Loop}
\State $d_{i,1} = f_1(d_{i,1} - w_1 d_{i,2})$
\EndFor
\end{algorithmic}
\end{algorithm}
\end{singlespacing}
%\If{$j <= 2$}
%\State $d_{i,j} = d_{i,j} r_j$
%\Else
%\State $d_{i,j} = f_j(d_{i,j} - r_i d_{i,j-1})$
%\EndIf

%First of all, the forward and backward passes of the hybrid Thomas-PCR algorithm are carried out based on the data structure such that the reduced sized system is obtained as shown in Figure \ref{fig:alg}.

After the decoupling phase of the algorithm is complete, the next step is solving the reduced systems.
As we are focusing on diagonally dominant systems, certain entries in the reduced tridiagonal system are significantly below the zero-machine value, and can be eliminated without any loss in numerical accuracy.
The entries that can be eliminated are indicated with red crosses in Figure \ref{fig:alg}.
Eliminating these entries simplifies the reduced systems further down to independent 2x2 systems as shown in red boxes in Figure \ref{fig:alg}, and each of these systems are coupled across an MPI boundary.
Although these systems are located in different ranks, solving them is trivial, and requires only local communication with first level neighbours.
The communication phase of the DistD2 algorithm therefore carries out an MPI communication with the two neighbouring ranks, and solves the $2\times2$ systems.
Single step MPI communication that involves only the previous and next neighbours is a significant advantage over existing strategies such as \cite{Laszlo2016}, \cite{Balogh2022}, and \cite{Kim2021} where solving the reduced tridiagonal systems require considerably more communications. 

Finally, the substitution phase of the DistD2 algorithm requires a simple algebraic substitution.
The first and last entries in each subdomain is solved in the communication phase, therefore, the remaining interior entries can be solved via Algorithm \ref{alg:CTDS_subs}.

\begin{singlespacing}
\begin{algorithm}[h!]
\caption{DistD2 - substitution phase.}\label{alg:CTDS_subs}
\begin{algorithmic}[1]
\For{$i = 1:SZ$}\Comment{Vectorised Loop}
\State $d_{i,1} = d^s_{i}$
\EndFor
\For{$j = 1:n$}
\For{$i = 1:SZ$}\Comment{Vectorised Loop}
\State $d_{i, j} = d_{i, j} - s^a_j d^s_i - s^c_j d^e_i$
\EndFor
\EndFor
\For{$i = 1:SZ$}\Comment{Vectorised Loop}
\State $d_{i,n} = d^e_{i}$
\EndFor
\end{algorithmic}
\end{algorithm}
\end{singlespacing}

The important point is that Algorithms \ref{alg:CTDS} and \ref{alg:CTDS_subs} consider a group of RHS arrays together as shown in Figure \ref{fig:pnclgroup}, and the inner $i$ loop over $SZ$ many RHS arrays indicates a vectorised execution on CPUs and thread level parallel execution on GPUs. %is responsible for ensuring the vectorisation.
On CPUs the vectorisation can be enabled via OpenMP \textit{simd} directive \cite{OpenMP}, and the kernel that solves the tridiagonal systems for a group of lines is called within a parallel for loop over all the groups in the domain.
On GPUs the thread level parallelism is achieved via executing the kernel with a thread dimension $SZ=32$, and a block dimension equal to the number of groups in the domain \cite{CUDA}.
Therefore, each thread is assigned a single RHS and the sequential operations across the $j$ loop do not block the concurrent execution of individual systems.

\begin{figure}[h!]
  \centering
  \input{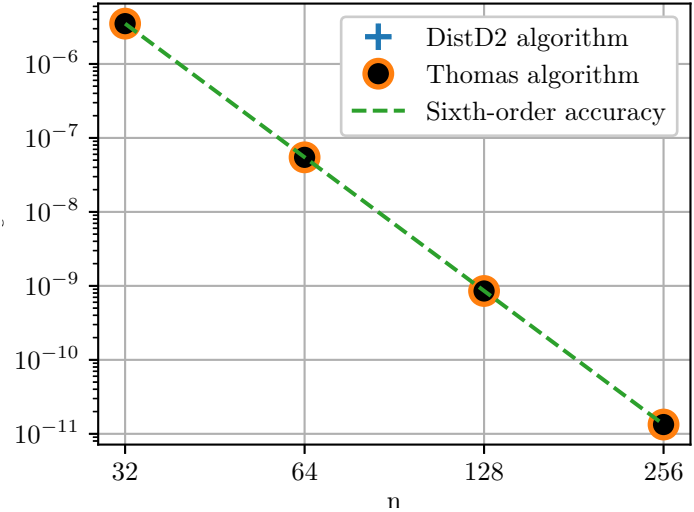}
  \caption{Order of accuracy of DistD2 algorithm compared against Thomas algorithm.}
  \label{fig:OoA-DistD2}
\end{figure}

Finally, we have carried out an order of accuracy analysis and compared DistD2 algorithm against Thomas algorithm to demonstrate that disregarding the zero-machine terms in the matrix do not reduce the accuracy of the algorithm.
The tests are carried out from 32 points up to 256 points for the first derivative using compact schemes in double precision with periodic boundary conditions such that boundary conditions do not impact the accuracy.
Periodic Thomas algorithm is executed on a single rank and DistD2 algorithm on 2 ranks.
The results are shown in Figure \ref{fig:OoA-DistD2} and demonstrates that DistD2 algorithm obtains identical results compared to the Thomas algorithm.

\section{Results and Performance}\label{sec:results}
In this section, we first focus on demonstrating the performance of the proposed algorithm on a single rank and at scale, and then assess the performance of the overall strategy on a multi-dimensional PDE.
The tests carried out in Sections 5.1 and 5.2 correspond to performing a numerical calculation that represents a single partial differential term in a multi-dimensional PDE discretised using compact finite difference schemes.
Section 5.3 on the other hand investigates solving a complete 3-dimensional PDE with all terms discretised using compact schemes and including the reorder operations that are required due to the specialised data structure when switching between spatial dimensions.
Therefore, this section presents the performance of the proposed strategy in a complete range starting from the underlying building blocks and up to a complete PDE.
\subsection{Single-rank Performance}
First, we investigate the single-rank performance both for the DistD2 algorithm and for the Thomas algorithm implemented using the proposed data structure on CPU and GPU architectures.
In order to provide a baseline for comparison, we worked out the data movements these algorithms require in terms of domain sized read, write, and read\&write operations when solving a batch of tridiagonal systems in a given domain.
Data movement requirements based on the algorithm are given in Table \ref{table:theoryBW} both for cached and standard implementations where cached indicates the intermediate states as the algorithm progresses are stored in cache, and standard indicates the intermediate states are not stored in cache, thus requiring additional data movements from the main memory.
\begin{table}[h]
\centering
\caption{Data movement requirements of all three algorithms examined in the manuscript including cached and standard implementations. Data movement requirements for the STREAM Copy and Scale benchmarks are also provided for comparison. All the numbers are normalised with respect to number of grid points.}
\begin{tabular}{l|r|rrr|rrr}
% &  & \multicolumn{2}{c}{Distributed}  & \multicolumn{2}{c}{Thomas} & \multicolumn{2}{c}{Thomas (periodic)}  \\
\multicolumn{2}{c}{} & \multicolumn{3}{c}{Standard} & \multicolumn{3}{c}{Cached} \\
               & FLOP     & R & W &R\&W & R & W & R\&W \\ \hline
%DistD2        & $\sim$18 & 1 & 1 & 2   & 1 & 1 & 1 \\
DistD2 decoupling phase & $\sim$14 & 1 & 1 & 1   & 1 & 1 & 0 \\
DistD2 substitution phase&  $\sim$4 & 0 & 0 & 1   & 0 & 0 & 1 \\
Thomas         & $\sim$14 & 1 & 1 & 1   & 1 & 1 & 0 \\
Periodic Thomas& $\sim$16 & 1 & 1 & 2   & 1 & 1 & 0 \\
STREAM \cite{STREAM} copy & - & 1 & 1 & 0   & - & - & -  \\
STREAM \cite{STREAM} scale (in-place)& $\sim$1 & 0 & 0 & 1 & - & - & -\\
\end{tabular}
\label{table:theoryBW}
\end{table}

\begin{figure}[h!]
  \centering
\begin{subfigure}{0.5\textwidth}
  \resizebox{\textwidth}{!}{\input{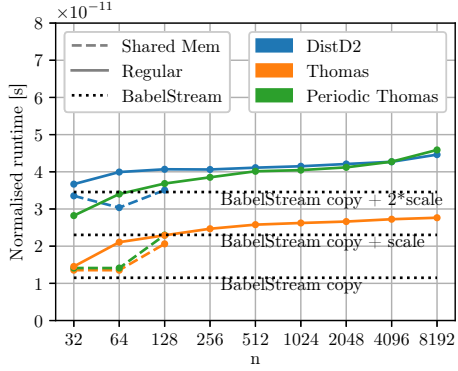}}
\caption{NVIDIA A100 (40GB HBM2)} \label{fig:perfGPU_a}
\end{subfigure}
\begin{subfigure}{0.49\textwidth}
  \resizebox{\textwidth}{!}{\input{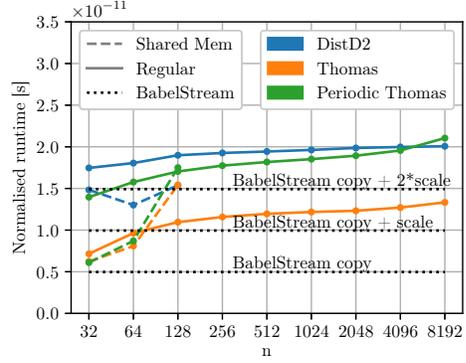}}
\caption{NVIDIA GH200 (96GB HBM3)} \label{fig:perfGPU_b}
\end{subfigure}
  \caption{Single rank performance of the DistD2 and Thomas algorithms on NVIDIA A100 (40GB HBM2)(a) and GH200 (96GB HBM3) (b) GPUs.}
  \label{fig:perfGPU}
\end{figure}

\begin{figure}[h!]
  \centering
\begin{subfigure}{0.5\textwidth}
  \resizebox{\textwidth}{!}{\input{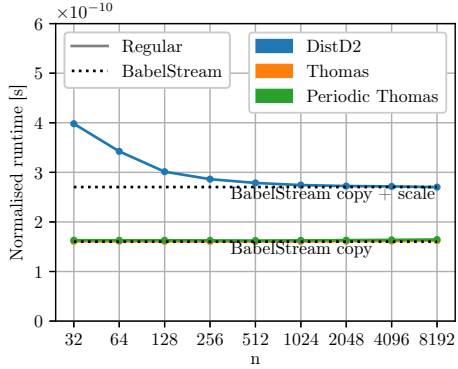}}
\caption{AMD EPYC 7742} \label{fig:perfCPU_a}
\end{subfigure}
\begin{subfigure}{0.49\textwidth}
  \resizebox{\textwidth}{!}{\input{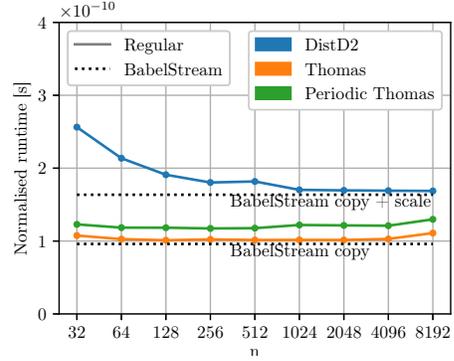}}
\caption{Intel Xeon Platinum 8460Y+} \label{fig:perfCPU_b}
\end{subfigure}
  \caption{Single rank performance of the DistD2 and Thomas algorithms on AMD EPYC 7742 (a) and Intel Xeon Planitum 8460Y+ (b) CPUs.}
  \label{fig:perfCPU}
\end{figure}

The FLOP requirements of these algorithms are very low as shown in Table \ref{table:theoryBW}, thus, we include copy and scale benchmarks from STREAM \cite{STREAM, STREAM2} that closely match the data movement requirements of the tridiagonal matrix solver algorithms.
The GPU and CPU benchmarks for all three tridiagonal matrix solver algorithms implemented based on the proposed data structure are shown in Figure \ref{fig:perfGPU} and Figure \ref{fig:perfCPU} respectively. 
The figures also include copy and scale benchmarks executed with BabelStream \cite{BabelStream} with a minor change in the source code such that the multiply (scale) benchmark carries out the operation in-place to reflect the data movement characteristics 
of DistD2 substitution phase
as shown in Table \ref{table:theoryBW}.
The benchmarks are carried out in a range of sizes of tridiagonal systems starting from 32 grid points per tridiagonal system up to 8192 grid points, and then the runtime is normalised with respect to the number of tridiagonal systems that are being solved in a 3D domain and the size of the individual tridiagonal systems.
The number of systems in a 3D domain is varied such that the total number of grid points are always constant and the problem require around 75\% of the available memory.
All the benchmarks are carried out in double precision and we observed less than 1\% variability in repeated tests.
Because the FLOP and data movement requirements scale linearly with the size of the tridiagonal systems and number of systems in a 3D domain, we anticipate a relatively constant normalised runtime across the range of tridiagonal system sizes.
The results shown in Figure \ref{fig:perfGPU} are consistent with our expectations.
First of all, due to the limited size of GPU shared memory, the shared memory implementation cannot go beyond a limited number of points per tridiagonal system.
However, in the range where the cache size is large enough to store the intermediate state, the performance of the kernels using shared memory are on par with the BabelStream copy benchmark for the periodic and non-periodic Thomas algorithms as indicated in Table \ref{table:theoryBW}.
Furthermore, the regular implementation of the DistD2 algorithm achieves a performance that matches the combination of one copy and two scale benchmarks, and the shared memory implementation of the DistD2 algorithm is on par with the combination of one copy and one scale benchmark as indicated in Table \ref{table:theoryBW}.
The regular implementations do not explicitly store the intermediate states in shared memory, instead rely on data to stay in L1 cache whenever possible.
Thus, we observe a better performance from the shared memory implementations.
In summary, as the trends in all three algorithms demonstrate up to n=8192, increasing the tridiagonal system size reduces the relative portion of the system that can be stored in cache, and a small reduction in performance is observed.
However, the implementations achieve a very high utilisation of the available bandwidth considering the required data movements. Furthermore, the runtimes are on par with BabelStream benchmarks that require equivalent levels of data movements.

Next, results in Figure \ref{fig:perfCPU} are also consistent with our data movement based performance model.
The per core private caches in modern CPUs are large enough for the sizes of tridiagonal systems that we are interested in.
For example, a group of tridiagonal systems with 4096 points and $SZ=4$ would only require around 256KiB of storage for the input and output arrays, and an additional 64-128KiB for the coefficient arrays depending on the algorithm in double precision.
The benchmarks in Figure \ref{fig:perfCPU} are carried out on AMD Zen2 and Intel Sapphire Rapids CPUs with 0.5MiB and 2MiB L2 cache per core, respectively.
Therefore, the intermediate states of the periodic and regular Thomas algorithms can be stored in cache, which makes the data movement requirements and performance on par with the BabelStream copy benchmark as shown in Figure \ref{fig:perfCPU}.
The DistD2 algorithm on the other hand consist of decoupling and substitution phases with MPI communications in between.
The decoupling phase writes the intermediate state back to main memory, and then the substitution phase reads the intermediate state from main memory.
This results in additional data movements and is reflected in the performance results.
The data movements of the decoupling phase of the algorithm are similar to the Thomas algorithm, however the substitution phase operates on the intermediate state in-place. 
Therefore, the performance of the DistD2 algorithm is roughly equivalent to combination of copy and scale benchmarks as shown in Figure \ref{fig:perfCPU}.
The higher normalised runtime on smaller tridiagonal systems on CPUs can be explained by %relatively higher workload due to 
halo data processing requiring relatively more work.

Finally, all three algorithms achieve a decent utilisation of the available bandwidth across a range of tridiagonal system sizes on a variety of architectures.
The bandwidth utilisations provided in Table \ref{table:BW} are based on the data movement requirements on each system for all three algorithms at 512 grid points per tridiagonal system.
\begin{table}[h]
\centering
\caption{Bandwidth utilisation on various hardware. The theoretical maximum available bandwidths are provided for each processor ('BW'). The achieved bandwidth is given as percentage of maximum available bandwidth ('achv') alongside the incurred data movement requirements per grid point ('req') based on the algorithm and hardware.}
\begin{tabular}{l r || r l | r l | r l}
 &  & \multicolumn{2}{c|}{DistD2}  & \multicolumn{2}{c|}{Thomas} & \multicolumn{2}{c}{Per. Thomas}  \\
 & BW & req & achv & req & achv & req & achv \\ \hline
AMD EPYC 7742 & 205 GB/s& 5 & 70.1\% & 3 & 73.1\% & 3 & 72.1\% \\
%Nvidia Grace CPU & 1000 GB/s & 4 & \textcolor{red}{?\%} & 2 & \textcolor{red}{?\%} & 2 & \textcolor{red}{?\%} \\
Intel Xeon 8460Y+ & 307 GB/s& 5 & 71.6\% & 3 & 77.0\% & 3 & 66.4\% \\
Nvidia A100 GPU & 1550 GB/s& 6 & 73.3\% & 4 & 77.4\% & 6 & 77.9\% \\
Nvidia V100 GPU & 900 GB/s& 6 & 68.7\% & 4 & 75.6\% & 6 & 73.6\% \\
Nvidia GH200 GPU & 4023 GB/s& 6 & 61.4\% & 4 & 66.6\% & 6 & 65.7\% \\
\end{tabular}
\label{table:BW}
\end{table}
It is important to note that AMD EPYC 7742 and Intel Xeon 8460Y+ use a write-allocate cache policy like many other modern x86 chips, which incurs an extra read operation to load the entries in cache before writing back to main memory \cite{STREAM}.
The other systems in Table \ref{table:BW} do not exhibit such behaviour, and this is reflected in the incurred data movements.
In summary, the results demonstrate that the proposed strategy is highly efficient and can achieve a very high utilisation of the available bandwidth near the practical maximum which is around 75-85\% as shown in \cite{BabelStream} with various STREAM benchmarks.

\subsection{Multi-rank Performance - Strong and Weak Scalability}
\iffalse
\begin{itemize}
    \item The communication is pseudo-local, only between the previous and next ranks.
    \item This enables an excellent weak and strong scalability as shown in \ref{fig:Scalability}. 
\end{itemize}
\fi
One of the key features of the proposed algorithm is that there is a significant reduction in the amount of communication between ranks compared to alternative approaches.
Solving a tridiagonal system in a distributed-memory environment with DistD2 algorithm requires only 2 sets of MPI communications.
The first one is to communicate the halo data which are data points located in a neighbouring rank that contribute to the RHS for the data points near the boundaries.
Depending on the stencil of the RHS, the depth of halo data is typically between 2 to 4. 
The second set of MPI communication is required in between the decoupling and substitution phases of the DistD2 algorithm.
The diagonally dominant tridiagonal systems are reduced down to 2$\times$2 systems across each MPI boundary with the decoupling phase of the DistD2 algorithm, and then the first and last grid points of each tridiagonal system are sent and received between the neighbouring ranks.
The DistD2 algorithm deviates from existing strategies at this step.
Most tridiagonal solvers use a PCR-type strategy to obtain the solution of the reduced system, or solve the reduced systems via a Thomas algorithm after gathering the reduced systems on a single rank.
Because we use compact schemes and always solve a diagonally dominant system, the reduced systems in our case are not coupled across the entire domain.
Instead, they are coupled across 2 individual ranks across an MPI boundary.
This is allowing us to carry out a single set of MPI communication and obtain the final result by carrying out the second phase of the DistD2 algorithm without any further MPI communications as in PCR strategies.

\begin{figure}[h!]
  \centering
  \input{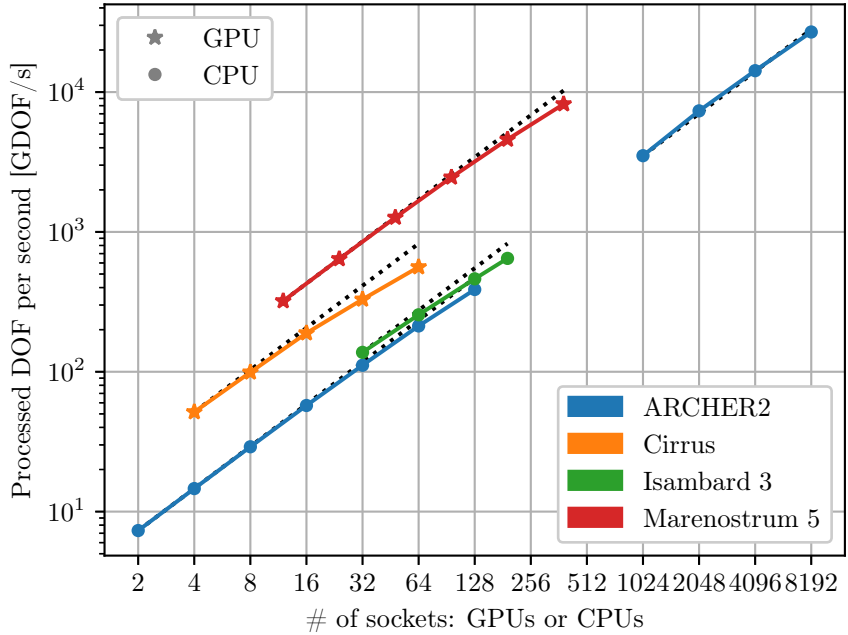}
  \caption{Strong scaling of DistD2 algorithm on ARCHER2 \cite{ARCHER2}, Cirrus \cite{Cirrus}, Isambard 3 \cite{Isambard3}, and MareNostrum 5 \cite{Marenostrum5} supercomputers.}
  \label{fig:Scalability}
\end{figure}

Figure \ref{fig:Scalability} demonstrate the strong scaling of our strategy on CPU and GPU based supercomputers.
In all benchmarks we used a sixth-order compact scheme to obtain the first derivative along a spatial dimension in double precision.
We start with a domain size that is big enough to use around 75\% of the available memory, which corresponds to 12.88 billion grid points on a single node on ARCHER2 (2x AMD EPYC 7742 CPU, 256 GiB DDR4 in total), 3.22 billion grid points on a single node on Cirrus (4x NVIDIA V100 GPU, each GPU with 16 GiB HBM), 11.95 billion grid points on a single node on Isambard 3 (2x NVIDIA Grace CPU, 240 GiB LPDDR5X in total), and 12.88 billion grid points on a single node on MareNostrum 5 (4x NVIDIA H100 GPU, each GPU with 64 GiB HBM2).
The domain is decomposed with a 1D decomposition strategy such that all the tridiagonal systems in the domain are divided into as many subdomains as there are sockets and distributed such that each socket operates on one subdomain from all of the tridiagonal systems in the domain. % and distributed across the MPI ranks.
The results demonstrate that the efficiency of strong scalability is 82.7\% on ARCHER2 from 2 to 128 CPUs and 95.8\% from 1024 to 8192 CPUs, 68.0\% on Cirrus from 4 to 64 GPUs, 78.3\% on Isambard from 32 to 192 CPUs, and 79.9\% on MareNostrum 5 from 12 to 384 GPUs.

\subsection{Application - 3D non-linear PDE}\label{sec:3D-PDE}
In this section we examine the performance of the proposed algorithm on a practical PDE in a 3D domain.
We apply compact scheme discretisation to the momentum transport equation, an important step when solving incompressible Navier-Stokes equations with fractional time stepping strategy \cite{Incompact3D}.
The momentum transport equation is often formulated in a skew-symmetric form using the continuity equation and this results in one additional term per equation.
The skew-symmetric formulation is useful for the stability when using collocated velocity fields for the incompressible Navier-Stokes equations.
The formulation of the momentum transport equation can be given in skew-symmetric form as,
\begin{equation}\label{eq:navstokes}
\frac{\partial \mathbf{u^*}}{\partial t}
=
-\frac{1}{2}
\big[
\nabla (\mathbf{u}\otimes\mathbf{u})
+ (\mathbf{u} \cdot \nabla)\mathbf{u}
\big]
+ \nu\nabla^2\mathbf{u}
\end{equation}
where $\mathbf{u}$ is the velocity, $\mathbf{u^*}$ denotes the intermediate field in fractional time stepping strategies, and $\nu$ is the kinematic viscosity.
Equation \ref{eq:navstokes} can be simplified and given in index notation as
\begin{equation}\label{eq:mom-comps}
\frac{\partial u_i^*}{\partial t}
= 
-\frac{1}{2}\Big(
u_j\frac{\partial u_i}{\partial x_j}
+
\frac{\partial u_j u_i}{\partial x_j}
\Big)
+ \nu\frac{\partial^2 u_i}{\partial x_j^2},
\end{equation}
where the terms operated on the same spatial direction are grouped together.

We use compact finite difference discretisation to evaluate the spatial derivative terms on the right hand side. 
The numerical characteristics of compact schemes have been studied in depth elsewhere \cite{Lele1992, Incompact3D},
therefore, in this section we evaluate the performance of the proposed strategy by focusing on the spatial discretisation and its implementation.

As discussed in Section \ref{sec:method}, the proposed data structure layout depends on the spatial direction of the derivative operations. 
Thus, in a multi-dimensional PDE there is a requirement to change the data layout for each direction.
In this benchmark we demonstrate that the data layout changes account for only a small percentage of the overall runtime, thus the overall strategy is efficient for solving 3D PDEs.

Furthermore, this equation system provides opportunities for fusing multiple operators into one to reduce the data movement requirements even further.
For example, $x$-directional derivatives in Equation \ref{eq:mom-comps} all require the $u$ field, thus, it is possible to reduce the overall data movements by implementing a dedicated kernel that reads the $u$ field once and solves for all three $x$-directional operators concurrently.
The abstract implementation based on DistD2 algorithm obtains the combination of three terms in the form of
\begin{equation}
RHS_j^{u_i} 
= 
-\frac{1}{2}\Big(
u_j\frac{\partial u_i}{\partial x_j}
+
\frac{\partial u_j u_i}{\partial x_j}
\Big)
+ \nu\frac{\partial^2 u_i}{\partial x_j^2},
\end{equation}
which requires 1 input field when $i\!=\!j$ and 2 input fields when $i\!\ne\!j$, and is executed three times per equation.
The operators in $y$- and $z$-directions require all three input fields to be reordered based on the specialised data structure explained in Section \ref{sec:method}, and then three solution fields in $y$- and $z$-directions need to be accumulated back into $x$-directional result fields.
Time integration can be carried out after this stage, however, as it is trivial and involves only vector additions we do not investigate its performance.

In order to assess the performance of the proposed strategy, we first identify the total amount of data movements our implementation requires including kernels carrying out data layout reorderings and result accumulations.
The data movement requirements are given in Table \ref{table:transport} normalised by the size of a scalar field on a given grid, per each kernel, specifying the number of times a kernel is called and the total data movement it contributes per full evaluation for two different platforms.
\begin{table}[h]
\centering
\caption{Data movement requirements for the transport equation on NVIDIA GPUs and x86 CPUs. A kernel fusion strategy is used on NVIDIA GPUs and a cache blocking strategy is used on x86 CPUs to minimise data movement requirements. $a+b$ format reflects the requirements in decoupling and substitution phases of DistD2 algorithm. '\#' indicates number of times a kernel is called, and 'T' indicates the incurred total data movement requirement based on the specific implementation and hardware.}
\begin{tabular}{l || r|rrr|r||rrr|r}
% &  & \multicolumn{2}{c}{Distributed}  & \multicolumn{2}{c}{Thomas} & \multicolumn{2}{c}{Thomas (periodic)}  \\
 \multicolumn{2}{c}{} & \multicolumn{4}{c}{NVIDIA GPU} & \multicolumn{4}{c}{x86 CPU} \\
                        & \# & R   & W   & R\&W & T  & R & W & R\&W & T \\ \hline
Abstract kernel ($i\!\ne\!j$) & 6 & 2+4 & 3+1 & 3+0   & 96 & 2+3 & 3+0 & 0+1 & 78 \\
Abstract kernel ($i\!=\!j$) & 3 & 1+4 & 3+1 & 3+0   & 45 & 1+3 & 3+0 & 0+1 & 36 \\
Reordering               & 6 & 1   & 1   & 0   & 12 & 1 & 1 & 0 & 18 \\
Accumulation             & 6 & 1   & 0   & 1   & 18 & 1 & 0 & 1 & 18 \\ \hline
\multicolumn{2}{l}{Total}   & \multicolumn{4}{r||}{171}  &   \multicolumn{4}{r}{150} \\
\end{tabular}
\label{table:transport}
\end{table}
The reordering operations that are introduced due to the specialised data structure account for only 7.0\% and 12.0\% of the total data movement requirement on GPUs and CPUs respectively.
Therefore, the impact of the additional operations required due to the specialised data structure are minimal when solving a multi-dimensional PDE based on the proposed strategy.

We have carried out the benchmarks on two different platforms at scale to demonstrate the robustness of the algorithm.
On both platforms we used a grid with $4096^3$ points resulting in 68.7 billion grid points in total and the domain is a square box with periodic boundary conditions.
The GPU benchmarks are carried out on Marenostrum 5 %where each node is equipped with 4x NVIDIA H100 [64GB, HBM2] GPUs \cite{Marenostrum5}, 
and the CPU benchmarks are carried out on ARCHER2. %where each node is equipped with 2x AMD EPYC 7742 CPUs with 256GB total memory \cite{ARCHER2}.
The available bandwidth on Marenostrum 5 is 4$\times$1600GB/s per node due to the HBM2 memory used in custom H100 GPUs, and 409GB/s
per node on ARCHER2 due to 8 memory channels with 3200MHz DDR4 setup.
We compare the actual runtimes from benchmarks and calculate the bandwidth utilisation by dividing the data movement requirements to runtimes.
The results are given in Table \ref{table:final}.
\begin{table}[h]
\centering
\caption{Runtime per step (t) and the sustained bandwidth utilisation as a ratio of available bandwidth (BW). MareNostrum 5 has 4x Nvidia H100 GPUs per node, and ARCHER2 has 2x AMD EPYC 7742 CPU per node.}
\begin{tabular}{l ||rr|rr}
% &  & \multicolumn{2}{c}{Distributed}  & \multicolumn{2}{c}{Thomas} & \multicolumn{2}{c}{Thomas (periodic)}  \\
 \multicolumn{1}{c}{} & \multicolumn{2}{c}{32 nodes} & \multicolumn{2}{c}{64 nodes} \\
               & t [s] & BW & t [s] & BW \\ \hline
MareNostrum 5 & 0.662 & 68.0\% & 0.337 & 66.9\% \\
ARCHER2     & 9.057 & 66.2\% & 4.568 & 65.7\% \\
\end{tabular}
\label{table:final}
\end{table}
Both systems have 256GB main memory per node, and the benchmarks use 87.5\% and 43.8\% of the available memory on 32 and 64 nodes respectively.
A 3D decomposition strategy is used due to the large size of the domain, and the decomposition on 32 nodes is $4\times4\times8$ on MareNostrum 5 such that there is 1 rank per GPU and $4\times8\times8$ on ARCHER2 such that there are 4 ranks per CPU to account for 4 NUMA zones.
The ratio of the runtime per step on MareNostrum 5 and ARCHER2 is around 13.6 and it is reasonable considering that a single MareNostrum 5 node has 15.65x more memory bandwidth but requires 14\% more data movements due to minor differences in implementation which brings expected speedup down to 13.73x.
The sustained bandwidth ranges from 66\% to 68\% of the theoretical peak, and the strong scalability from 32 nodes to 64 nodes is 98.3\% on MareNostrum 5 and 99.1\% on ARCHER2.

\section{Conclusion}\label{sec:conclusion}
Solving a batch of tridiagonal systems in distributed-memory environments is a challenging problem.
First of all, most existing strategies require more communication as the number of ranks increase, preventing a linear strong scaling.
This is particularly a problem with large scale supercomputers with thousands of CPUs or GPUs.
We have developed a novel algorithm that uses the diagonal dominance in certain tridiagonal systems to minimise the communication requirements, limiting it to be between neighbouring regions in the computational domain.
Additionally, we proposed a specialised data structure to improve the performance by using data localities and data continuity.
As a result, we minimised the data movements, achieved a near peak bandwidth utilisation, and vectorisation on CPUs and thread level parallelism on GPUs.
We also demonstrated the strong scaling on up to 8192 AMD EPYC 7742 CPUs on ARCHER2, and up to 384 NVIDIA H100 GPUs on MareNostrum 5.
Finally, we provided a practical application using compact finite difference schemes and solved a 3D PDE to demonstrate the robustness of the algorithm.
The performance analysis indicates that DistD2 algorithm sustains around 66\% of theoretical peak bandwidth on ARCHER2, and 67\% of theoretical peak bandwidth on MareNostrum 5 at scale when solving a practical 3D PDE.

\section{Acknowledgements}
The authors would like to thank the Engineering and Physical Research Council (EPSRC) for their support via grant number EP/W026686/1, the Edinburgh Parallel Computing Centre (EPCC) for their support via grants ARCHER2 CPUeCSE10-1 and GPUeCSE0110, EuroHPC for their support via grant number EHPC-DEV-2024D06-021 providing development access to MareNostrum5, the Bristol Supercomputing Centre (BriCS) for their support via early access calls to Isambard 3 and Isambard-AI, and the UK Turbulence Consortium for providing access to ARCHER2 and Cirrus via grant number EP/X035484/1.
For the purpose of open access, the author has applied a Creative Commons Attribution (CC BY) licence to any Author Accepted Manuscript version arising from this submission.



%% References

%\section*{References}
\bibliographystyle{elsarticle-num}
\bibliography{CPC_TDS}






% \end{thebibliography}

\end{document}